\documentclass[10pt,conference]{IEEEtran}
\usepackage[letterpaper, left = 0.68in, right = 0.68in, top = 0.75in, bottom = 1.05in]{geometry}

\usepackage{tikz}
\usetikzlibrary{spy}
\usepackage{color, colortbl}
\usepackage{pgfplots}
\pgfplotsset{compat=newest}
\usepackage{multicol}
\usepackage{circuitikz}
\usepackage[nolist]{acronym}
\usepgfplotslibrary{groupplots}
\usetikzlibrary{shapes,math}
\usepackage{siunitx}  
\usepackage{graphicx}
\usepgfplotslibrary{patchplots}
\usepackage{amsmath,amssymb}
\usepackage{subfig}
 \usepackage{float}
\usepackage{cite}
\usepackage{amsmath}
\usetikzlibrary{angles, quotes,calc,patterns}
\usepgfplotslibrary{fillbetween}

\usepackage{ifthen}
\usepackage{algorithm}
\usepackage[noend]{algpseudocode}
\usetikzlibrary{matrix,calc}
\usepackage{trfsigns}
\usepackage[font=small]{caption}

\usepackage{textcomp} %
\usepackage{nicefrac}
\usepackage{multirow}
\usepackage{booktabs}
\usepackage[shortcuts,acronym]{glossaries}
\usepackage{bm}
\usepackage{lipsum}
\usepackage{soul} %
\usepackage{balance}

\definecolor{uni_apfelgruen}{cmyk}{.5, 0, 1, 0}
\definecolor{uni_mittelblau}{cmyk}{1, 0.4, 0, 0}
\definecolor{uni_bulletblau}{RGB}{49,99,183}
\definecolor{uni_gelb}{cmyk}{0, 0.1, 1, 0}
\definecolor{uni_rot}{cmyk}{0, 1, 1, 0}
\definecolor{uniblauHell}{RGB}{0,190,255}
\definecolor{uniblauDunkel}{RGB}{0,65,145}
\definecolor{unigrau}{RGB}{51,51,51}
\definecolor{darkgray176}{RGB}{176,176,176}
\definecolor{lavenderplot}{RGB}{191,148,228}
\definecolor{coralplot}{RGB}{255,127,80}
\definecolor{cyanplot}{RGB}{37,219,168}

\newacronym{adc}{ADC}{analog-to-digital converter}
\newacronym{aoa}{AoA}{angle-of-arrival}
\newacronym{aod}{AoD}{angle-of-departure}
\newacronym{tc-ta85}{TC-TA85}{R\&S\textsuperscript{\small\textregistered}TC-TA85CP cross-polarized Vivaldi test antenna}
\newacronym{areg}{AREG}{R\&S\textsuperscript{\small\textregistered}AREG800A automotive radar echo generator}
\newacronym{awgn}{AWGN}{additive white Gaussian noise}
\newacronym{dac}{DAC}{digital-to-analog converter}
\newacronym{cdf}{CDF}{cumulative distribution function}
\newacronym{cacfar}{CA-CFAR}{cell averaging constant false alarm rate}
\newacronym{cfar}{CFAR}{constant false alarm rate}
\newacronym{cir}{CIR}{channel impulse response}
\newacronym{cphd}{CPHD}{cardinalized probability hypothesis density}
\newacronym{crap}{CRAP}{clutter removal with acquisitions under phase noise}
\newacronym{csi}{CSI}{channel state information}
\newacronym{ctf}{CTF}{channel transfer function}
\newacronym{dft}{DFT}{discrete Fourier transform}
\newacronym{dl}{DL}{deep learning}
\newacronym{doli}{DL}{downlink}
\newacronym{eca}{ECA}{extensive cancellation algorithm}
\newacronym{eca-c}{ECA-C}{extensive cancellation algorithm by subcarrier}
\newacronym{esprit}{ESPRIT}{estimation of signal parameters via rotational invariance techniques}
\newacronym{fe44s}{FE44S}{R\&S\textsuperscript{\small\textregistered}FE44S external frontend}
\newacronym{fisst}{FISST}{finite set statistics}
\newacronym{gmix}{GM}{Gaussian mixture}
\newacronym{gnn}{GNN}{global nearest neighbor}
\newacronym{isac}{ISAC}{integrated sensing and communication}
\newacronym{jpdaf}{JPDAF}{joint probabilistic data association filter}
\newacronym{kf}{KF}{Kalman filter}
\newacronym{lmf}{LMF}{location management function}
\newacronym{mae}{MAE}{mean absolute error}
\newacronym{mht}{MHT}{multiple hypothesis tracking}
\newacronym{mimo}{MIMO}{multiple-input and multiple-output}
\newacronym{ml}{ML}{maximum likelihood}
\newacronym{mtt}{MTT}{multi-target tracking}
\newacronym{naf}{NAF}{normalized angular frequency}
\newacronym{nf}{NF}{noise figure}
\newacronym{ofdm}{OFDM}{orthogonal frequency-division multiplexing}
\newacronym{phd}{PHD}{probability hypothesis density}
\newacronym{poc}{PoC}{proof-of-concept}
\newacronym{psf}{PSF}{point spread function}
\newacronym{radar}{radar}{radio detection and ranging}
\newacronym{rfs}{RFS}{random finite set}
\newacronym{rcs}{RCS}{radar cross-section}
\newacronym{rmse}{RMSE}{root mean square error}
\newacronym{ru}{RU}{radio unit}
\newacronym{rx}{Rx}{receiver}
\newacronym{sap}{SAP}{sensing access point}
\newacronym{semf}{SeMF}{sensing management function}
\newacronym{slam}{SLAM}{simultaneous localization and mapping}
\newacronym{snr}{SNR}{signal-to-noise ratio}
\newacronym{tdd}{TDD}{time division duplex}
\newacronym{tx}{Tx}{transmitter}
\newacronym{upli}{UL}{uplink}
\newacronym{ula}{ULA}{uniform linear array}

\title{Experimental Demonstration of Multi-Target Tracking in Integrated Sensing and Communication}

\author{%
\IEEEauthorblockN{Maximilian Bauhofer\IEEEauthorrefmark{1}, Marcus Henninger\IEEEauthorrefmark{2}, Meik Kottkamp\IEEEauthorrefmark{3}, Lucas Giroto\IEEEauthorrefmark{2}, Philip Grill\IEEEauthorrefmark{3}}%
\IEEEauthorblockN{Alexander Felix\IEEEauthorrefmark{2}, Thorsten Wild\IEEEauthorrefmark{2}, Stephan ten Brink\IEEEauthorrefmark{1}, and Silvio Mandelli\IEEEauthorrefmark{2}}%

\IEEEauthorblockA{ %
    \IEEEauthorrefmark{1}University of Stuttgart, Stuttgart %
    \IEEEauthorrefmark{2}Nokia Bell Labs, Stuttgart %
    \IEEEauthorrefmark{3}Rohde \& Schwarz, Munich, Germany%
}
\IEEEauthorblockA{E-mail: bauhofer@inue.uni-stuttgart.de}
}

\newcommand\blfootnote[1]{%
  \begingroup
  \renewcommand\thefootnote{}\footnote{#1}%
  \addtocounter{footnote}{-1}%
  \endgroup
}

\begin{document}

\pgfplotsset{
    colormap={jet_inue}{
        rgb=(1, 1, 1)
        rgb=(0.99804, 0.99804, 0.99905)
        rgb=(0.99609, 0.99609, 0.99813)
        rgb=(0.99413, 0.99413, 0.99725)
        rgb=(0.99217, 0.99217, 0.99639)
        rgb=(0.99022, 0.99022, 0.99557)
        rgb=(0.98826, 0.98826, 0.99477)
        rgb=(0.9863, 0.9863, 0.99401)
        rgb=(0.98434, 0.98434, 0.99327)
        rgb=(0.98239, 0.98239, 0.99257)
        rgb=(0.98043, 0.98043, 0.9919)
        rgb=(0.97847, 0.97847, 0.99125)
        rgb=(0.97652, 0.97652, 0.99064)
        rgb=(0.97456, 0.97456, 0.99006)
        rgb=(0.9726, 0.9726, 0.98951)
        rgb=(0.97065, 0.97065, 0.98899)
        rgb=(0.96869, 0.96869, 0.9885)
        rgb=(0.96673, 0.96673, 0.98804)
        rgb=(0.96477, 0.96477, 0.98762)
        rgb=(0.96282, 0.96282, 0.98722)
        rgb=(0.96086, 0.96086, 0.98685)
        rgb=(0.9589, 0.9589, 0.98652)
        rgb=(0.95695, 0.95695, 0.98621)
        rgb=(0.95499, 0.95499, 0.98593)
        rgb=(0.95303, 0.95303, 0.98569)
        rgb=(0.95108, 0.95108, 0.98548)
        rgb=(0.94912, 0.94912, 0.98529)
        rgb=(0.94716, 0.94716, 0.98514)
        rgb=(0.94521, 0.94521, 0.98502)
        rgb=(0.94325, 0.94325, 0.98493)
        rgb=(0.94129, 0.94129, 0.98486)
        rgb=(0.93933, 0.93933, 0.98483)
        rgb=(0.93738, 0.93738, 0.98483)
        rgb=(0.93542, 0.93542, 0.98486)
        rgb=(0.93346, 0.93346, 0.98493)
        rgb=(0.93151, 0.93151, 0.98502)
        rgb=(0.92955, 0.92955, 0.98514)
        rgb=(0.92759, 0.92759, 0.98529)
        rgb=(0.92564, 0.92564, 0.98548)
        rgb=(0.92368, 0.92368, 0.98569)
        rgb=(0.92172, 0.92172, 0.98593)
        rgb=(0.91977, 0.91977, 0.98621)
        rgb=(0.91781, 0.91781, 0.98652)
        rgb=(0.91585, 0.91585, 0.98685)
        rgb=(0.91389, 0.91389, 0.98722)
        rgb=(0.91194, 0.91194, 0.98762)
        rgb=(0.90998, 0.90998, 0.98804)
        rgb=(0.90802, 0.90802, 0.9885)
        rgb=(0.90607, 0.90607, 0.98899)
        rgb=(0.90411, 0.90411, 0.98951)
        rgb=(0.90215, 0.90215, 0.99006)
        rgb=(0.9002, 0.9002, 0.99064)
        rgb=(0.89824, 0.89824, 0.99125)
        rgb=(0.89628, 0.89628, 0.9919)
        rgb=(0.89432, 0.89432, 0.99257)
        rgb=(0.89237, 0.89237, 0.99327)
        rgb=(0.89041, 0.89041, 0.99401)
        rgb=(0.88845, 0.88845, 0.99477)
        rgb=(0.8865, 0.8865, 0.99557)
        rgb=(0.88454, 0.88454, 0.99639)
        rgb=(0.88258, 0.88258, 0.99725)
        rgb=(0.88063, 0.88063, 0.99813)
        rgb=(0.87867, 0.87867, 0.99905)
        rgb=(0.87671, 0.87671, 1)
        rgb=(0.87476, 0.87573, 1)
        rgb=(0.8728, 0.87479, 1)
        rgb=(0.87084, 0.87387, 1)
        rgb=(0.86888, 0.87298, 1)
        rgb=(0.86693, 0.87213, 1)
        rgb=(0.86497, 0.8713, 1)
        rgb=(0.86301, 0.87051, 1)
        rgb=(0.86106, 0.86974, 1)
        rgb=(0.8591, 0.86901, 1)
        rgb=(0.85714, 0.8683, 1)
        rgb=(0.85519, 0.86763, 1)
        rgb=(0.85323, 0.86699, 1)
        rgb=(0.85127, 0.86638, 1)
        rgb=(0.84932, 0.8658, 1)
        rgb=(0.84736, 0.86525, 1)
        rgb=(0.8454, 0.86473, 1)
        rgb=(0.84344, 0.86424, 1)
        rgb=(0.84149, 0.86378, 1)
        rgb=(0.83953, 0.86335, 1)
        rgb=(0.83757, 0.86295, 1)
        rgb=(0.83562, 0.86259, 1)
        rgb=(0.83366, 0.86225, 1)
        rgb=(0.8317, 0.86194, 1)
        rgb=(0.82975, 0.86167, 1)
        rgb=(0.82779, 0.86142, 1)
        rgb=(0.82583, 0.86121, 1)
        rgb=(0.82387, 0.86103, 1)
        rgb=(0.82192, 0.86087, 1)
        rgb=(0.81996, 0.86075, 1)
        rgb=(0.818, 0.86066, 1)
        rgb=(0.81605, 0.8606, 1)
        rgb=(0.81409, 0.86057, 1)
        rgb=(0.81213, 0.86057, 1)
        rgb=(0.81018, 0.8606, 1)
        rgb=(0.80822, 0.86066, 1)
        rgb=(0.80626, 0.86075, 1)
        rgb=(0.80431, 0.86087, 1)
        rgb=(0.80235, 0.86103, 1)
        rgb=(0.80039, 0.86121, 1)
        rgb=(0.79843, 0.86142, 1)
        rgb=(0.79648, 0.86167, 1)
        rgb=(0.79452, 0.86194, 1)
        rgb=(0.79256, 0.86225, 1)
        rgb=(0.79061, 0.86259, 1)
        rgb=(0.78865, 0.86295, 1)
        rgb=(0.78669, 0.86335, 1)
        rgb=(0.78474, 0.86378, 1)
        rgb=(0.78278, 0.86424, 1)
        rgb=(0.78082, 0.86473, 1)
        rgb=(0.77886, 0.86525, 1)
        rgb=(0.77691, 0.8658, 1)
        rgb=(0.77495, 0.86638, 1)
        rgb=(0.77299, 0.86699, 1)
        rgb=(0.77104, 0.86763, 1)
        rgb=(0.76908, 0.8683, 1)
        rgb=(0.76712, 0.86901, 1)
        rgb=(0.76517, 0.86974, 1)
        rgb=(0.76321, 0.87051, 1)
        rgb=(0.76125, 0.8713, 1)
        rgb=(0.7593, 0.87213, 1)
        rgb=(0.75734, 0.87298, 1)
        rgb=(0.75538, 0.87387, 1)
        rgb=(0.75342, 0.87479, 1)
        rgb=(0.75147, 0.87573, 1)
        rgb=(0.74951, 0.87671, 1)
        rgb=(0.74755, 0.87772, 1)
        rgb=(0.7456, 0.87876, 1)
        rgb=(0.74364, 0.87983, 1)
        rgb=(0.74168, 0.88093, 1)
        rgb=(0.73973, 0.88206, 1)
        rgb=(0.73777, 0.88323, 1)
        rgb=(0.73581, 0.88442, 1)
        rgb=(0.73386, 0.88564, 1)
        rgb=(0.7319, 0.88689, 1)
        rgb=(0.72994, 0.88818, 1)
        rgb=(0.72798, 0.88949, 1)
        rgb=(0.72603, 0.89084, 1)
        rgb=(0.72407, 0.89222, 1)
        rgb=(0.72211, 0.89362, 1)
        rgb=(0.72016, 0.89506, 1)
        rgb=(0.7182, 0.89653, 1)
        rgb=(0.71624, 0.89802, 1)
        rgb=(0.71429, 0.89955, 1)
        rgb=(0.71233, 0.90111, 1)
        rgb=(0.71037, 0.9027, 1)
        rgb=(0.70841, 0.90432, 1)
        rgb=(0.70646, 0.90597, 1)
        rgb=(0.7045, 0.90766, 1)
        rgb=(0.70254, 0.90937, 1)
        rgb=(0.70059, 0.91111, 1)
        rgb=(0.69863, 0.91289, 1)
        rgb=(0.69667, 0.91469, 1)
        rgb=(0.69472, 0.91652, 1)
        rgb=(0.69276, 0.91839, 1)
        rgb=(0.6908, 0.92028, 1)
        rgb=(0.68885, 0.92221, 1)
        rgb=(0.68689, 0.92417, 1)
        rgb=(0.68493, 0.92616, 1)
        rgb=(0.68297, 0.92817, 1)
        rgb=(0.68102, 0.93022, 1)
        rgb=(0.67906, 0.9323, 1)
        rgb=(0.6771, 0.93441, 1)
        rgb=(0.67515, 0.93655, 1)
        rgb=(0.67319, 0.93872, 1)
        rgb=(0.67123, 0.94092, 1)
        rgb=(0.66928, 0.94316, 1)
        rgb=(0.66732, 0.94542, 1)
        rgb=(0.66536, 0.94771, 1)
        rgb=(0.66341, 0.95004, 1)
        rgb=(0.66145, 0.95239, 1)
        rgb=(0.65949, 0.95478, 1)
        rgb=(0.65753, 0.95719, 1)
        rgb=(0.65558, 0.95964, 1)
        rgb=(0.65362, 0.96211, 1)
        rgb=(0.65166, 0.96462, 1)
        rgb=(0.64971, 0.96716, 1)
        rgb=(0.64775, 0.96973, 1)
        rgb=(0.64579, 0.97233, 1)
        rgb=(0.64384, 0.97496, 1)
        rgb=(0.64188, 0.97762, 1)
        rgb=(0.63992, 0.98031, 1)
        rgb=(0.63796, 0.98303, 1)
        rgb=(0.63601, 0.98578, 1)
        rgb=(0.63405, 0.98856, 1)
        rgb=(0.63209, 0.99138, 1)
        rgb=(0.63014, 0.99422, 1)
        rgb=(0.62818, 0.9971, 1)
        rgb=(0.62622, 1, 1)
        rgb=(0.6272, 1, 0.99706)
        rgb=(0.62821, 1, 0.9941)
        rgb=(0.62925, 1, 0.9911)
        rgb=(0.63032, 1, 0.98807)
        rgb=(0.63142, 1, 0.98502)
        rgb=(0.63255, 1, 0.98193)
        rgb=(0.63371, 1, 0.97881)
        rgb=(0.63491, 1, 0.97566)
        rgb=(0.63613, 1, 0.97248)
        rgb=(0.63738, 1, 0.96927)
        rgb=(0.63867, 1, 0.96603)
        rgb=(0.63998, 1, 0.96276)
        rgb=(0.64133, 1, 0.95945)
        rgb=(0.6427, 1, 0.95612)
        rgb=(0.64411, 1, 0.95276)
        rgb=(0.64555, 1, 0.94936)
        rgb=(0.64702, 1, 0.94594)
        rgb=(0.64851, 1, 0.94248)
        rgb=(0.65004, 1, 0.939)
        rgb=(0.6516, 1, 0.93548)
        rgb=(0.65319, 1, 0.93193)
        rgb=(0.65481, 1, 0.92836)
        rgb=(0.65646, 1, 0.92475)
        rgb=(0.65815, 1, 0.92111)
        rgb=(0.65986, 1, 0.91744)
        rgb=(0.6616, 1, 0.91374)
        rgb=(0.66337, 1, 0.91001)
        rgb=(0.66518, 1, 0.90625)
        rgb=(0.66701, 1, 0.90246)
        rgb=(0.66888, 1, 0.89864)
        rgb=(0.67077, 1, 0.89478)
        rgb=(0.6727, 1, 0.8909)
        rgb=(0.67466, 1, 0.88699)
        rgb=(0.67665, 1, 0.88304)
        rgb=(0.67866, 1, 0.87907)
        rgb=(0.68071, 1, 0.87506)
        rgb=(0.68279, 1, 0.87102)
        rgb=(0.6849, 1, 0.86696)
        rgb=(0.68704, 1, 0.86286)
        rgb=(0.68921, 1, 0.85873)
        rgb=(0.69141, 1, 0.85457)
        rgb=(0.69365, 1, 0.85039)
        rgb=(0.69591, 1, 0.84617)
        rgb=(0.6982, 1, 0.84192)
        rgb=(0.70053, 1, 0.83763)
        rgb=(0.70288, 1, 0.83332)
        rgb=(0.70527, 1, 0.82898)
        rgb=(0.70768, 1, 0.82461)
        rgb=(0.71013, 1, 0.82021)
        rgb=(0.7126, 1, 0.81577)
        rgb=(0.71511, 1, 0.81131)
        rgb=(0.71765, 1, 0.80681)
        rgb=(0.72022, 1, 0.80229)
        rgb=(0.72282, 1, 0.79773)
        rgb=(0.72545, 1, 0.79314)
        rgb=(0.72811, 1, 0.78853)
        rgb=(0.7308, 1, 0.78388)
        rgb=(0.73352, 1, 0.7792)
        rgb=(0.73627, 1, 0.77449)
        rgb=(0.73905, 1, 0.76975)
        rgb=(0.74187, 1, 0.76498)
        rgb=(0.74471, 1, 0.76018)
        rgb=(0.74758, 1, 0.75535)
        rgb=(0.75049, 1, 0.75049)
        rgb=(0.75342, 1, 0.7456)
        rgb=(0.75639, 1, 0.74067)
        rgb=(0.75939, 1, 0.73572)
        rgb=(0.76241, 1, 0.73074)
        rgb=(0.76547, 1, 0.72572)
        rgb=(0.76856, 1, 0.72068)
        rgb=(0.77168, 1, 0.7156)
        rgb=(0.77483, 1, 0.71049)
        rgb=(0.77801, 1, 0.70536)
        rgb=(0.78122, 1, 0.70019)
        rgb=(0.78446, 1, 0.69499)
        rgb=(0.78773, 1, 0.68976)
        rgb=(0.79103, 1, 0.6845)
        rgb=(0.79437, 1, 0.67921)
        rgb=(0.79773, 1, 0.67389)
        rgb=(0.80113, 1, 0.66854)
        rgb=(0.80455, 1, 0.66316)
        rgb=(0.80801, 1, 0.65775)
        rgb=(0.81149, 1, 0.65231)
        rgb=(0.81501, 1, 0.64683)
        rgb=(0.81855, 1, 0.64133)
        rgb=(0.82213, 1, 0.63579)
        rgb=(0.82574, 1, 0.63023)
        rgb=(0.82938, 1, 0.62463)
        rgb=(0.83305, 1, 0.61901)
        rgb=(0.83675, 1, 0.61335)
        rgb=(0.84048, 1, 0.60766)
        rgb=(0.84424, 1, 0.60194)
        rgb=(0.84803, 1, 0.5962)
        rgb=(0.85185, 1, 0.59042)
        rgb=(0.85571, 1, 0.58461)
        rgb=(0.85959, 1, 0.57877)
        rgb=(0.8635, 1, 0.5729)
        rgb=(0.86745, 1, 0.56699)
        rgb=(0.87142, 1, 0.56106)
        rgb=(0.87543, 1, 0.5551)
        rgb=(0.87946, 1, 0.54911)
        rgb=(0.88353, 1, 0.54308)
        rgb=(0.88763, 1, 0.53703)
        rgb=(0.89176, 1, 0.53094)
        rgb=(0.89591, 1, 0.52483)
        rgb=(0.9001, 1, 0.51868)
        rgb=(0.90432, 1, 0.51251)
        rgb=(0.90857, 1, 0.5063)
        rgb=(0.91285, 1, 0.50006)
        rgb=(0.91717, 1, 0.49379)
        rgb=(0.92151, 1, 0.48749)
        rgb=(0.92588, 1, 0.48116)
        rgb=(0.93028, 1, 0.4748)
        rgb=(0.93472, 1, 0.46841)
        rgb=(0.93918, 1, 0.46199)
        rgb=(0.94368, 1, 0.45554)
        rgb=(0.9482, 1, 0.44906)
        rgb=(0.95276, 1, 0.44255)
        rgb=(0.95734, 1, 0.436)
        rgb=(0.96196, 1, 0.42943)
        rgb=(0.96661, 1, 0.42282)
        rgb=(0.97129, 1, 0.41619)
        rgb=(0.976, 1, 0.40952)
        rgb=(0.98074, 1, 0.40283)
        rgb=(0.98551, 1, 0.3961)
        rgb=(0.99031, 1, 0.38934)
        rgb=(0.99514, 1, 0.38255)
        rgb=(1, 1, 0.37573)
        rgb=(1, 0.99511, 0.37378)
        rgb=(1, 0.99018, 0.37182)
        rgb=(1, 0.98523, 0.36986)
        rgb=(1, 0.98025, 0.36791)
        rgb=(1, 0.97523, 0.36595)
        rgb=(1, 0.97019, 0.36399)
        rgb=(1, 0.96511, 0.36204)
        rgb=(1, 0.96, 0.36008)
        rgb=(1, 0.95487, 0.35812)
        rgb=(1, 0.9497, 0.35616)
        rgb=(1, 0.9445, 0.35421)
        rgb=(1, 0.93927, 0.35225)
        rgb=(1, 0.93401, 0.35029)
        rgb=(1, 0.92872, 0.34834)
        rgb=(1, 0.9234, 0.34638)
        rgb=(1, 0.91805, 0.34442)
        rgb=(1, 0.91267, 0.34247)
        rgb=(1, 0.90726, 0.34051)
        rgb=(1, 0.90182, 0.33855)
        rgb=(1, 0.89634, 0.33659)
        rgb=(1, 0.89084, 0.33464)
        rgb=(1, 0.8853, 0.33268)
        rgb=(1, 0.87974, 0.33072)
        rgb=(1, 0.87414, 0.32877)
        rgb=(1, 0.86852, 0.32681)
        rgb=(1, 0.86286, 0.32485)
        rgb=(1, 0.85717, 0.3229)
        rgb=(1, 0.85146, 0.32094)
        rgb=(1, 0.84571, 0.31898)
        rgb=(1, 0.83993, 0.31703)
        rgb=(1, 0.83412, 0.31507)
        rgb=(1, 0.82828, 0.31311)
        rgb=(1, 0.82241, 0.31115)
        rgb=(1, 0.81651, 0.3092)
        rgb=(1, 0.81057, 0.30724)
        rgb=(1, 0.80461, 0.30528)
        rgb=(1, 0.79862, 0.30333)
        rgb=(1, 0.79259, 0.30137)
        rgb=(1, 0.78654, 0.29941)
        rgb=(1, 0.78045, 0.29746)
        rgb=(1, 0.77434, 0.2955)
        rgb=(1, 0.76819, 0.29354)
        rgb=(1, 0.76202, 0.29159)
        rgb=(1, 0.75581, 0.28963)
        rgb=(1, 0.74957, 0.28767)
        rgb=(1, 0.7433, 0.28571)
        rgb=(1, 0.737, 0.28376)
        rgb=(1, 0.73068, 0.2818)
        rgb=(1, 0.72432, 0.27984)
        rgb=(1, 0.71792, 0.27789)
        rgb=(1, 0.7115, 0.27593)
        rgb=(1, 0.70505, 0.27397)
        rgb=(1, 0.69857, 0.27202)
        rgb=(1, 0.69206, 0.27006)
        rgb=(1, 0.68551, 0.2681)
        rgb=(1, 0.67894, 0.26614)
        rgb=(1, 0.67233, 0.26419)
        rgb=(1, 0.6657, 0.26223)
        rgb=(1, 0.65903, 0.26027)
        rgb=(1, 0.65234, 0.25832)
        rgb=(1, 0.64561, 0.25636)
        rgb=(1, 0.63885, 0.2544)
        rgb=(1, 0.63206, 0.25245)
        rgb=(1, 0.62524, 0.25049)
        rgb=(1, 0.6184, 0.24853)
        rgb=(1, 0.61152, 0.24658)
        rgb=(1, 0.6046, 0.24462)
        rgb=(1, 0.59766, 0.24266)
        rgb=(1, 0.59069, 0.2407)
        rgb=(1, 0.58369, 0.23875)
        rgb=(1, 0.57666, 0.23679)
        rgb=(1, 0.56959, 0.23483)
        rgb=(1, 0.5625, 0.23288)
        rgb=(1, 0.55538, 0.23092)
        rgb=(1, 0.54822, 0.22896)
        rgb=(1, 0.54103, 0.22701)
        rgb=(1, 0.53382, 0.22505)
        rgb=(1, 0.52657, 0.22309)
        rgb=(1, 0.51929, 0.22114)
        rgb=(1, 0.51199, 0.21918)
        rgb=(1, 0.50465, 0.21722)
        rgb=(1, 0.49728, 0.21526)
        rgb=(1, 0.48988, 0.21331)
        rgb=(1, 0.48245, 0.21135)
        rgb=(1, 0.47499, 0.20939)
        rgb=(1, 0.4675, 0.20744)
        rgb=(1, 0.45997, 0.20548)
        rgb=(1, 0.45242, 0.20352)
        rgb=(1, 0.44484, 0.20157)
        rgb=(1, 0.43722, 0.19961)
        rgb=(1, 0.42958, 0.19765)
        rgb=(1, 0.42191, 0.19569)
        rgb=(1, 0.4142, 0.19374)
        rgb=(1, 0.40646, 0.19178)
        rgb=(1, 0.3987, 0.18982)
        rgb=(1, 0.3909, 0.18787)
        rgb=(1, 0.38307, 0.18591)
        rgb=(1, 0.37521, 0.18395)
        rgb=(1, 0.36733, 0.182)
        rgb=(1, 0.35941, 0.18004)
        rgb=(1, 0.35146, 0.17808)
        rgb=(1, 0.34347, 0.17613)
        rgb=(1, 0.33546, 0.17417)
        rgb=(1, 0.32742, 0.17221)
        rgb=(1, 0.31935, 0.17025)
        rgb=(1, 0.31125, 0.1683)
        rgb=(1, 0.30311, 0.16634)
        rgb=(1, 0.29495, 0.16438)
        rgb=(1, 0.28675, 0.16243)
        rgb=(1, 0.27853, 0.16047)
        rgb=(1, 0.27027, 0.15851)
        rgb=(1, 0.26199, 0.15656)
        rgb=(1, 0.25367, 0.1546)
        rgb=(1, 0.24532, 0.15264)
        rgb=(1, 0.23694, 0.15068)
        rgb=(1, 0.22853, 0.14873)
        rgb=(1, 0.2201, 0.14677)
        rgb=(1, 0.21163, 0.14481)
        rgb=(1, 0.20312, 0.14286)
        rgb=(1, 0.19459, 0.1409)
        rgb=(1, 0.18603, 0.13894)
        rgb=(1, 0.17744, 0.13699)
        rgb=(1, 0.16882, 0.13503)
        rgb=(1, 0.16016, 0.13307)
        rgb=(1, 0.15148, 0.13112)
        rgb=(1, 0.14277, 0.12916)
        rgb=(1, 0.13402, 0.1272)
        rgb=(1, 0.12524, 0.12524)
        rgb=(0.99315, 0.12329, 0.12329)
        rgb=(0.98627, 0.12133, 0.12133)
        rgb=(0.97936, 0.11937, 0.11937)
        rgb=(0.97242, 0.11742, 0.11742)
        rgb=(0.96545, 0.11546, 0.11546)
        rgb=(0.95845, 0.1135, 0.1135)
        rgb=(0.95141, 0.11155, 0.11155)
        rgb=(0.94435, 0.10959, 0.10959)
        rgb=(0.93726, 0.10763, 0.10763)
        rgb=(0.93013, 0.10568, 0.10568)
        rgb=(0.92298, 0.10372, 0.10372)
        rgb=(0.91579, 0.10176, 0.10176)
        rgb=(0.90857, 0.099804, 0.099804)
        rgb=(0.90133, 0.097847, 0.097847)
        rgb=(0.89405, 0.09589, 0.09589)
        rgb=(0.88674, 0.093933, 0.093933)
        rgb=(0.8794, 0.091977, 0.091977)
        rgb=(0.87203, 0.09002, 0.09002)
        rgb=(0.86463, 0.088063, 0.088063)
        rgb=(0.8572, 0.086106, 0.086106)
        rgb=(0.84974, 0.084149, 0.084149)
        rgb=(0.84225, 0.082192, 0.082192)
        rgb=(0.83473, 0.080235, 0.080235)
        rgb=(0.82718, 0.078278, 0.078278)
        rgb=(0.81959, 0.076321, 0.076321)
        rgb=(0.81198, 0.074364, 0.074364)
        rgb=(0.80434, 0.072407, 0.072407)
        rgb=(0.79666, 0.07045, 0.07045)
        rgb=(0.78896, 0.068493, 0.068493)
        rgb=(0.78122, 0.066536, 0.066536)
        rgb=(0.77345, 0.064579, 0.064579)
        rgb=(0.76566, 0.062622, 0.062622)
        rgb=(0.75783, 0.060665, 0.060665)
        rgb=(0.74997, 0.058708, 0.058708)
        rgb=(0.74208, 0.056751, 0.056751)
        rgb=(0.73416, 0.054795, 0.054795)
        rgb=(0.72621, 0.052838, 0.052838)
        rgb=(0.71823, 0.050881, 0.050881)
        rgb=(0.71022, 0.048924, 0.048924)
        rgb=(0.70218, 0.046967, 0.046967)
        rgb=(0.6941, 0.04501, 0.04501)
        rgb=(0.686, 0.043053, 0.043053)
        rgb=(0.67787, 0.041096, 0.041096)
        rgb=(0.6697, 0.039139, 0.039139)
        rgb=(0.66151, 0.037182, 0.037182)
        rgb=(0.65328, 0.035225, 0.035225)
        rgb=(0.64503, 0.033268, 0.033268)
        rgb=(0.63674, 0.031311, 0.031311)
        rgb=(0.62842, 0.029354, 0.029354)
        rgb=(0.62008, 0.027397, 0.027397)
        rgb=(0.6117, 0.02544, 0.02544)
        rgb=(0.60329, 0.023483, 0.023483)
        rgb=(0.59485, 0.021526, 0.021526)
        rgb=(0.58638, 0.019569, 0.019569)
        rgb=(0.57788, 0.017613, 0.017613)
        rgb=(0.56935, 0.015656, 0.015656)
        rgb=(0.56079, 0.013699, 0.013699)
        rgb=(0.5522, 0.011742, 0.011742)
        rgb=(0.54357, 0.0097847, 0.0097847)
        rgb=(0.53492, 0.0078278, 0.0078278)
        rgb=(0.52624, 0.0058708, 0.0058708)
        rgb=(0.51752, 0.0039139, 0.0039139)
        rgb=(0.50878, 0.0019569, 0.0019569)
        rgb=(0.5, 0, 0)
    }
}

\maketitle
\thispagestyle{empty}
\pagestyle{empty}

\begin{abstract}
For a wide range of envisioned \gls{isac} use cases, it is necessary to incorporate tracking techniques into cellular communication systems.
While numerous \gls{mtt} algorithms exist, they have not yet been applied to real-world \gls{isac}, with its challenges such as clutter and non-optimal hardware with design emphasis on communication instead of sensing.
In this work, we showcase \gls{mtt} based on the \gls{phd} filter in the range and radial speed domain.
The measurements are taken with a 5G compliant \gls{isac} proof-of-concept in a real factory environment, where the pedestrian-like targets are generated by a \acrshort{radar} target emulator.
We detail the complete pipeline, from measurement acquisition to evaluation, with a focus on the post-processing of the raw captured data and the tracking itself.
Our end-to-end evaluation and comparison to simulations show good \gls{mtt} performance with mean absolute ranging error \textless\,1.5\,m and detection rates \textgreater\,91\% for realistic but challenging scenarios.
\end{abstract}

\blfootnote{This work has been submitted to the IEEE for possible publication.
Copyright may be transferred without notice, after which this version may no longer be accessible.}

\acresetall  %

\glsresetall

\vspace{-4mm}
\section{Introduction}\label{sec:introduction}
Starting with 6G, future cellular networks are envisioned to incorporate \gls{isac}~\cite{mandelli_isacsurvey,liu_isac}.
Hereby, the wireless devices constitute a sensor network, providing information about the surrounding world.
Examples for this are \acrshort{radar}-like detection of targets, mobile handover prediction, \gls{slam}, and digital twins, with the first being the most imminent.

There is already a plethora of works investigating how, for example, a base station can extract the relevant range, radial speed (from Doppler shift) and angular information from commercial communication hardware and signals~\cite{braun_diss}.
The raw sensor data needs further processing to make it usable for higher layer applications.
For this, tracking is commonly employed.
Such techniques not only provide smoother tracks by incorporating past target states, but can also counter missed detections and false alarms.
The latter are particularly amplified due to communication-centric system design not tailored to sensing in terms of, e.g., hardware and signal structures.

For the single target case this is well established, typically using various flavors of conventional Bayesian filtering, such as the Kalman filter~\cite{richards2010principles}.
For the more general and challenging case of \gls{mtt}, there exist algorithms under the families of \gls{gnn}, \glspl{jpdaf}, \gls{mht}, and \glspl{rfs}~\cite{vo2015_mtt}.
While the first two require an estimate of the number of targets and need to be extended to enable target birth and death, the latter two include this natively.
All techniques but \gls{rfs} are grounded in a suboptimal split of data association with subsequent single target filtering.
\gls{rfs}, on the other hand, skips this on the basis of a rigorous extension of the Bayes filter to the \gls{mtt} case.
At this point, machine learning does not seem to be practical for an end-to-end tracking system due to limited training data, but can excel by replacing specific parts such as target classification~\cite{chong2021mltracking}.

In \gls{isac}, tracking has not yet been well researched despite conditions like clutter richness and suboptimal hardware compared to conventional \acrshort{radar} applications necessitating such techniques.
Single target tracking has been investigated in purely simulative studies for drone detection~\cite{jiang2024uav}, WiFi~\cite{tai2024target}, inverse synthetic aperture \acrshort{radar} (ISAR)~\cite{gui2024isar}, multistatic setups~\cite{yuan2025kalman}, and extended target using machine learning techniques~\cite{wang2025deep}.
The authors of~\cite{liesegang2025scalableintegratedsensingcommunications} include multi-targets in a simulated multistatic environment, but focus on the trade-off between communication and sensing.
The work in \cite{Liu_2023} describes outdoor multi-static multi-target measurements with 5G NR compliant base stations, but does not employ an actual tracking algorithm and evaluates performance on a snapshot-to-snapshot basis.

In this work, we apply a multi-target \gls{rfs} filter to real-world \gls{isac} measurements.
Hereby, we focus on the delay-Doppler domain, i.e., we track unlabeled targets based on their range and radial speed.
Our monostatic setup measures an indoor pedestrian scenario based on target emulation with real hardware.
The benefit of this intermediate step towards measurements with true physical targets are that perfect ground truth is precisely available, enabling comparison to pure simulations.
Furthermore, all hardware effects are present and the clutter from the real environment for a realistic performance.\newline
Our contributions are as follows:
\begin{enumerate}
    \item to the best of our knowledge, we showcase the world's first \gls{isac} tracking based on measurements with a communication system
    \item we provide the detailed pipeline, from acquisition to tracking result using a state-of-the-art \acrshort{phd} filter,
    \item we validate our implementation end-to-end in a real-world industrial scenario with multiple emulated targets using real hardware.
\end{enumerate}

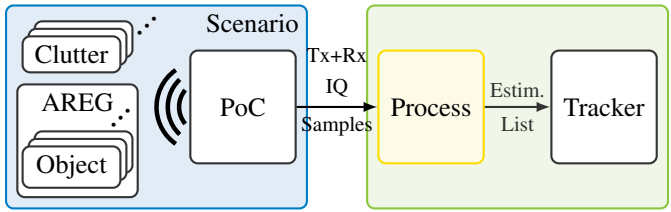
\begin{figure}[t!]
	\centering
  	\begin{tikzpicture}

    \def\height{2.5cm}
    \def\width{1.4cm}
    
    \node[draw=uni_mittelblau,fill=uni_mittelblau!10, thick, rounded corners=.1cm, minimum height=\height+0.2cm,minimum width=4.0cm,align=left] (onl) at (2.05,1) {};
    \node[draw=uni_apfelgruen,fill=uni_apfelgruen!10, thick, rounded corners=.1cm, minimum height=\height+0.2cm,minimum width=4.0cm,align=left] (offl) at (4.5+2.2+0.15,1) {};

    \node[draw=unigrau,fill=white, thick, rounded corners=.1cm, minimum height=\height-1cm,minimum width=\width,align=left] (poc) at (3.2,1) {\acrshort{poc}};

    \node[draw=uni_gelb,fill=uni_gelb!10, thick, rounded corners=.1cm, minimum height=\height-1cm,minimum width=\width,align=left] (proc) at (5.7,1) {Process};

    \node[draw=unigrau,fill=white, thick, rounded corners=.1cm, minimum height=\height-1cm,minimum width=\width,align=left] (track) at (8.0,1) {Tracker};

    \node[draw=unigrau,fill=white, thick, rounded corners=.1cm, minimum width=\width-0.1cm,align=left] (clutter1) at (1+0.08,1.8+0.1) {\phantom{Clutter}};
    \node[draw=unigrau,fill=white, thick, rounded corners=.1cm, minimum width=\width-0.1cm,align=left] (clutter2) at (1,1.8) {\phantom{Clutter}};
    \node[draw=unigrau,fill=white, thick, rounded corners=.1cm, minimum width=\width-0.1cm,align=left] (clutter3) at (1-0.08,1.8-0.1) {Clutter};
    
    \node[draw=unigrau,fill=white, thick, rounded corners=.1cm, minimum height=1.5cm,minimum width=\width+0.2cm,align=left,] (taremul) at (1.0,0.55) {};
    \node[draw=unigrau,fill=white, thick, rounded corners=.1cm, minimum width=\width-0.1cm,align=left] (obj1) at (1+0.08,0.3+0.1) {\phantom{Target}};
    \node[draw=unigrau,fill=white, thick, rounded corners=.1cm, minimum width=\width-0.1cm,align=left] (obj2) at (1,0.3) {\phantom{Target}};
    \node[draw=unigrau,fill=white, thick, rounded corners=.1cm, minimum width=\width-0.1cm,align=left] (obj3) at (1-0.08,0.3-0.1) {Target};

    \node[rotate=45] at ($(clutter1.east) + (0.20,0.2)$) {\small{\textbf{\dots}}};
    \node[rotate=45] at ($(obj1.east) + (-0.15,0.48)$) {\small{\textbf{\dots}}};
    \node[anchor=north] at (taremul.north) {\acrshort{areg}};
    
    \draw[-latex, thick, black] (poc.east) -- (proc.west) node[pos=0.5, above, sloped, align=center] {\footnotesize{\acrshort{tx}+\acrshort{rx}}\\\footnotesize{IQ}}  node[pos=0.5, below, sloped] {\footnotesize{Samples}};
    \draw[-latex,thick, unigrau] (proc.east) -- (track.west) node[pos=0.5, above, sloped] {\footnotesize{Estim.}}  node[pos=0.5, below, sloped] {\footnotesize{List}};

    \draw[thick, line width=0.6mm] (2.95-0.25,1) +(90+50:0.35) arc (90+50:270-50:0.35);  %
    \draw[thick, line width=0.6mm] (2.95-0.25,1) +(90+43:0.5) arc (90+43:270-43:0.5);
    \draw[thick, line width=0.6mm] (2.9-0.25,1) +(90+35:0.6) arc (90+35:270-35:0.6);

    \node[anchor=north east] at (onl.north east) {Scenario};

\end{tikzpicture}
    \vspace*{-4mm}
	\caption{Block diagram of the complete tracker setup.
    The 'online' part is shown inside the blue rectangle, with the \acrshort{poc} measuring the scenario of emulated targets and clutter.
    The green rectangle comprises the offline processing chain and tracking of targets.}
	\label{fig:setup:blockdiagram}
    \vspace*{-2mm}
\end{figure}

\begin{figure}[ht]
	\centering
  	\input{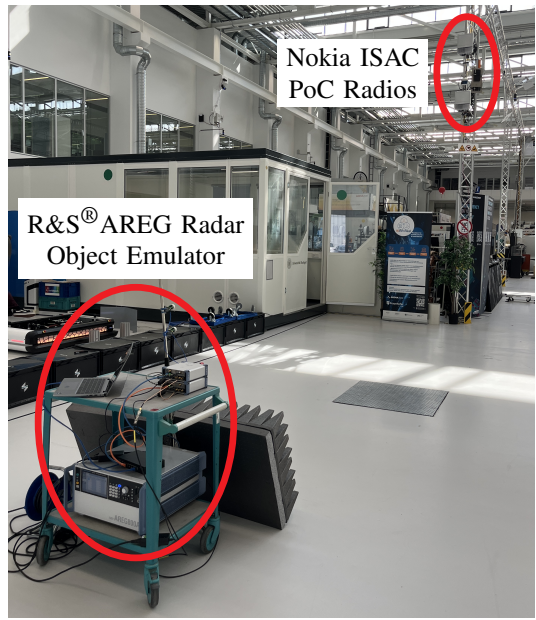}
    \caption{Image of the measurement setup in the ARENA2036.
    In the foreground, the \acrshort{radar} target emulator; in the background, the quasi-monostatic 5G compliant \acrshort{isac} \acrshort{poc} \acrshortpl{ru} for transmission and reception.}
	\label{fig:setup:arena_sceanrio}
    \vspace*{-8mm}
\end{figure}

\section{System Setup}\label{sec:setup}
We run our tracker based on real-world measurements.
Our system setup consists of two independent parts as depicted in Fig.~\ref{fig:setup:blockdiagram}.
The online part in blue is the actual acquisition of measurement data in the real environment with multiple emulated scenarios and further described below.
The offline part in green comprises all the post-processing and the actual tracker involved, which is described in the following chapters.
While the system could be implemented in a live demonstrator, we keep it separated for better tracker investigations.

Fig.~\ref{fig:setup:arena_sceanrio} shows the setup for measurement acquisition.
The Nokia \gls{isac} \gls{poc} radiates a 5G-NR compliant \gls{ofdm} signal towards the \gls{areg}.
This device captures the signal, modulates target states onto it, and re-radiates it towards the \gls{isac} \gls{poc}, where it is captured for further processing.

\subsection{ISAC PoC}\label{subsec:setup:poc}
The Nokia \gls{isac} \gls{poc} consists of two identical, unaltered, commercially available FR2 \glspl{ru} operating at \qty{27.6}{\giga\hertz}.
The upper \gls{ru} transmits 5G compliant \gls{ofdm} signals, the lower one receives the reflections from the environment.
The radios are configured in a quasi-monostatic setup, mounted at a height of ca. \qty{5.6}{\meter} with \qty{1}{\meter} vertical spacing.
This separation is sufficient to create enough isolation and circumvent the full-duplex issue of self-interference from sensing transmitter to receiver.
However, it is small enough that the system can be treated as approximately monostatic.
The radios have analog beam steering capabilities, which are set to a fixed beam that is tilted downwards toward the investigation area here, effectively collapsing the angular domain. %
The maximum output power of the transmitting \gls{ru} with \qty{55}{\dB\milli\relax} EIRP is adjusted to comply with health and indoor regulatory requirements as well as not to saturate the target emulator \glspl{adc}.
The transmission, reception, and storing of the \gls{ofdm} frame IQ samples is managed from a central server monitoring the information flow on the eCPRI interface of both radios.
For a more extensive description of the \gls{poc}, please refer to~\cite{wild20236g}.

\subsection{\Acrshort{radar} Target Emulator}\label{subsec:setup:areg}
\begin{figure}[t]    
    \centering
    \includegraphics[width=0.82\columnwidth]{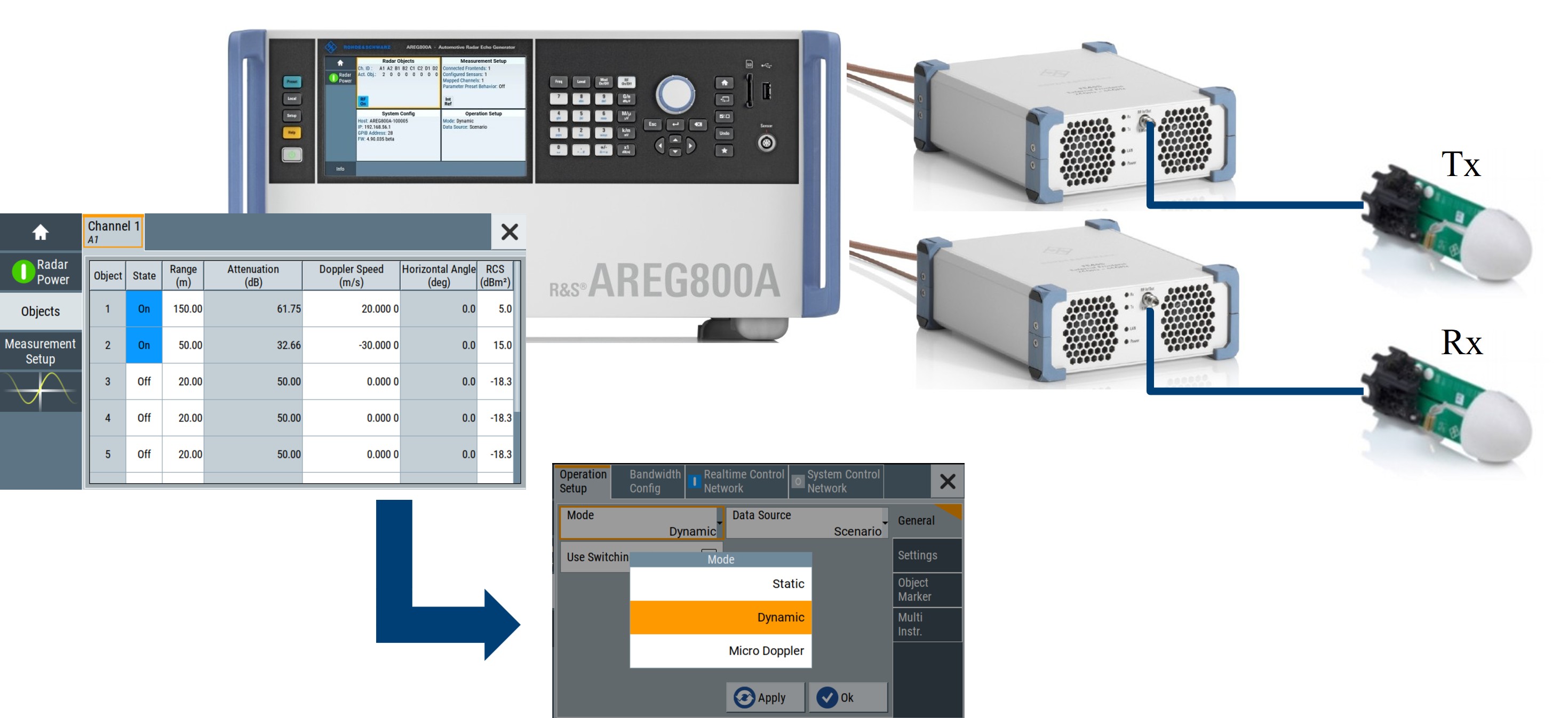}
    \caption{Illustration of the setup enabling static and dynamic target emulation including FR2 frontends and broadband antennas.}
    \label{fig:setup:areg_setup_illu}
    \vspace*{-3mm}
\end{figure}

Verifying \gls{isac} sensing performance is often based on physical targets reflecting high power, such as metallic plates, balls or real persons, bicycles, cars, which are placed in or moved within the scenario of interest. Testing based on physical targets is generally limited in terms of the number of targets available and the parameter range -- here distance or range and radial speed from Doppler shift -- that can be covered. Furthermore, reliability and reproducibility for comparison of different solutions present a challenge. As an alternative, a test instrument enables sensing performance verification based on emulated targets. In this paper, we applied target emulation based on the \gls{areg}. Its combination with \acrlong{fe44s} frequency up/down conversion enables target emulation at the frequency of \qty{27.6}{\giga\hertz}. We used \acrlongpl{tc-ta85} to enable over-the-air measurements of the test scenarios, see Fig.~\ref{fig:setup:areg_setup_illu}. Both static and dynamic target emulation is possible on the basis of point targets. The configurable parameters for each target are distance (range), radial speed, and \gls{rcs}. 

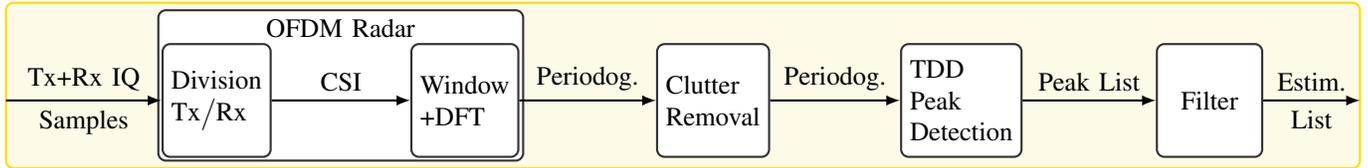
\begin{figure*}[t]
	\centering
  	\begin{tikzpicture}

    \def\height{2cm}
    \def\width{1.4cm}
    
    \pgfdeclarelayer{layer0}
    \pgfdeclarelayer{layer1}
    \pgfdeclarelayer{layer2}
    \pgfdeclarelayer{layer3}
    \pgfdeclarelayer{layer4}
    \pgfsetlayers{main,layer0,layer1,layer2,layer3,layer4}
    
    \begin{pgfonlayer}{layer0}

        \node[draw=uni_gelb,fill=uni_gelb!10, thick, rounded corners=.1cm, minimum height=\height+0.2cm,minimum width=18cm,align=left] (proc) at (9,1) {};
        \node[draw=unigrau,fill=white, thick, rounded corners=.1cm, minimum height=\height,minimum width=4.85cm,align=left, ] (rad) at (4.45,1) {};
        
    \end{pgfonlayer}
    
    \begin{pgfonlayer}{layer1}
        \node[draw=unigrau,fill=white, thick, rounded corners=.1cm, minimum height=\height-0.5cm,minimum width=\width,align=left] (div) at (2.8,0.8) {Division\\\acrshort{tx}\big/\acrshort{rx}};
        \node[draw=unigrau,fill=white, thick, rounded corners=.1cm, minimum height=\height-0.5cm,minimum width=\width,align=left] (per) at (2.8+3.3,0.8) {Window\\+\acrshort{dft}};
        \node[draw=unigrau,fill=white, thick, rounded corners=.1cm, minimum height=\height-0.5cm,minimum width=\width,align=left] (clutter) at (2.8+3.3*2,0.8) {Clutter\\Removal};
        \node[draw=unigrau,fill=white, thick, rounded corners=.1cm, minimum height=\height-0.5cm,minimum width=\width,align=left] (tdd) at (2.8+3.3*3,0.8) {\acrshort{tdd}\\Peak\\Detection};
        \node[draw=unigrau,fill=white, thick, rounded corners=.1cm, minimum height=\height-0.5cm,minimum width=\width,align=left] (filt) at (18-2,0.8) {Filter};

    \end{pgfonlayer}
            
    \begin{pgfonlayer}{layer4}

        \draw[-latex, thick, black] (0,0.8) -- (div.west) node[pos=0.5, above, sloped] {\acrshort{tx}+\acrshort{rx} IQ} node[pos=0.5, below, sloped] {Samples};
        \draw[-latex, thick, black] (div.east) -- (per.west) node[pos=0.5, above, sloped] {\acrshort{csi}};
        \draw[-latex, thick, black] (per.east) -- (clutter.west) node[pos=0.5, above, sloped] {Periodog.};
        \draw[-latex, thick, black] (clutter.east) -- (tdd.west) node[pos=0.5, above, sloped] {Periodog.};
        \draw[-latex, thick, black] (tdd.east) -- (filt.west) node[pos=0.5, above, sloped] {Peak List};

        \draw[-latex, thick, black] (filt.east) -- (18,0.8) node[pos=0.5, above, sloped] {Estim.} node[pos=0.5, below, sloped] {List};
        
        \node[anchor=north] at (rad.north) {\acrshort{ofdm} Radar};
    \end{pgfonlayer}

\end{tikzpicture}
    \vspace*{-4mm}
	\caption{Block diagram of the processing pipeline.
    It handles \gls{ofdm} \acrshort{radar} signal processing, management of scenario and hardware effects and returns a list of estimates.}
	\label{fig:pipeline:blockdiagram}
    \vspace*{-5mm}
\end{figure*}

The availability of the perfect ground truth based on the instrument settings allows reliable and reproducible evaluation of the tracker algorithms applied.
On the other side, emulated point scatterers do not resemble full-sized targets, which provides a limitation for target classification.
The main benefit of this setup is the safe creation of different complex test scenarios under the realistic abstraction from target effects.
On top of the target emulation we need to cope with all the \gls{poc} related hardware effects and nonlinearities including the complete real-world but non-perfect indoor scenario with its inherent clutter.
For cleaner measurements, we tried to keep the scenario free of obvious clutter such as people and targets in direct vicinity.

\section{Processing Pipeline}\label{sec:pipeline}

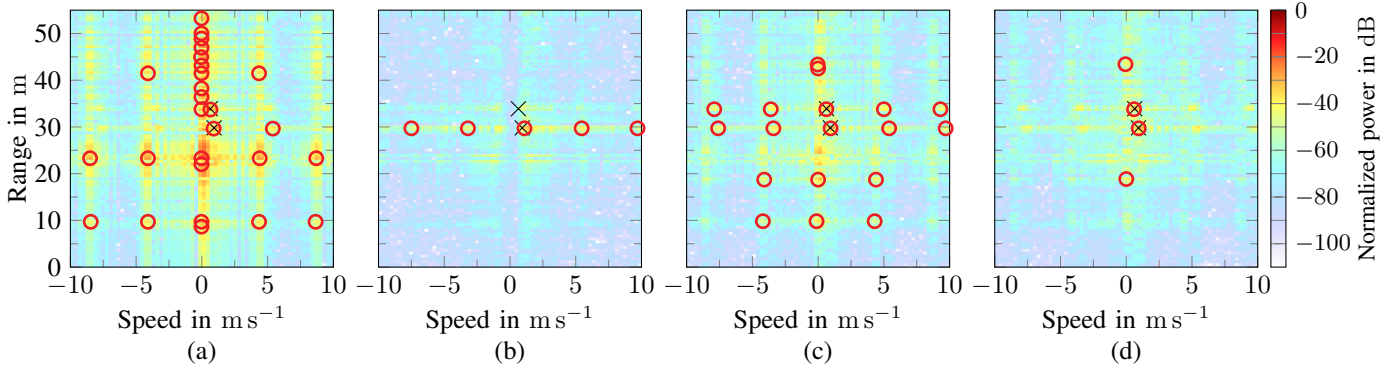
\begin{figure*}[t]
	\centering
  	\begin{tikzpicture}
    \centering
    \begin{groupplot}[
        group style={group name=periodograms, group size=4 by 1, vertical sep=0mm, horizontal sep=6mm},
        width=0.28\textwidth, height=5cm, %
        axis on top,
        xmin=-10, xmax=10, xlabel={Speed in \unit{\meter\per\second}}, xlabel near ticks, xtick={-10,-5,...,10}, xlabel style={yshift=1mm}, minor x tick num=1,
        ymin=0, ymax=55, ytick={0,10,...,50}, yticklabels={,,}, ylabel={},
        ylabel near ticks, ylabel style={yshift=-2mm}, minor y tick num=1,
        ]

        \nextgroupplot[ylabel={Range in \unit{\meter}}, yticklabels={0,10,...,50}, xlabel={Speed in \unit{\meter\per\second}\\(a)}, xlabel style={align=center}]

            \addplot graphics [xmin=-10, xmax=10, ymin=-0.3022, ymax=55.3022] {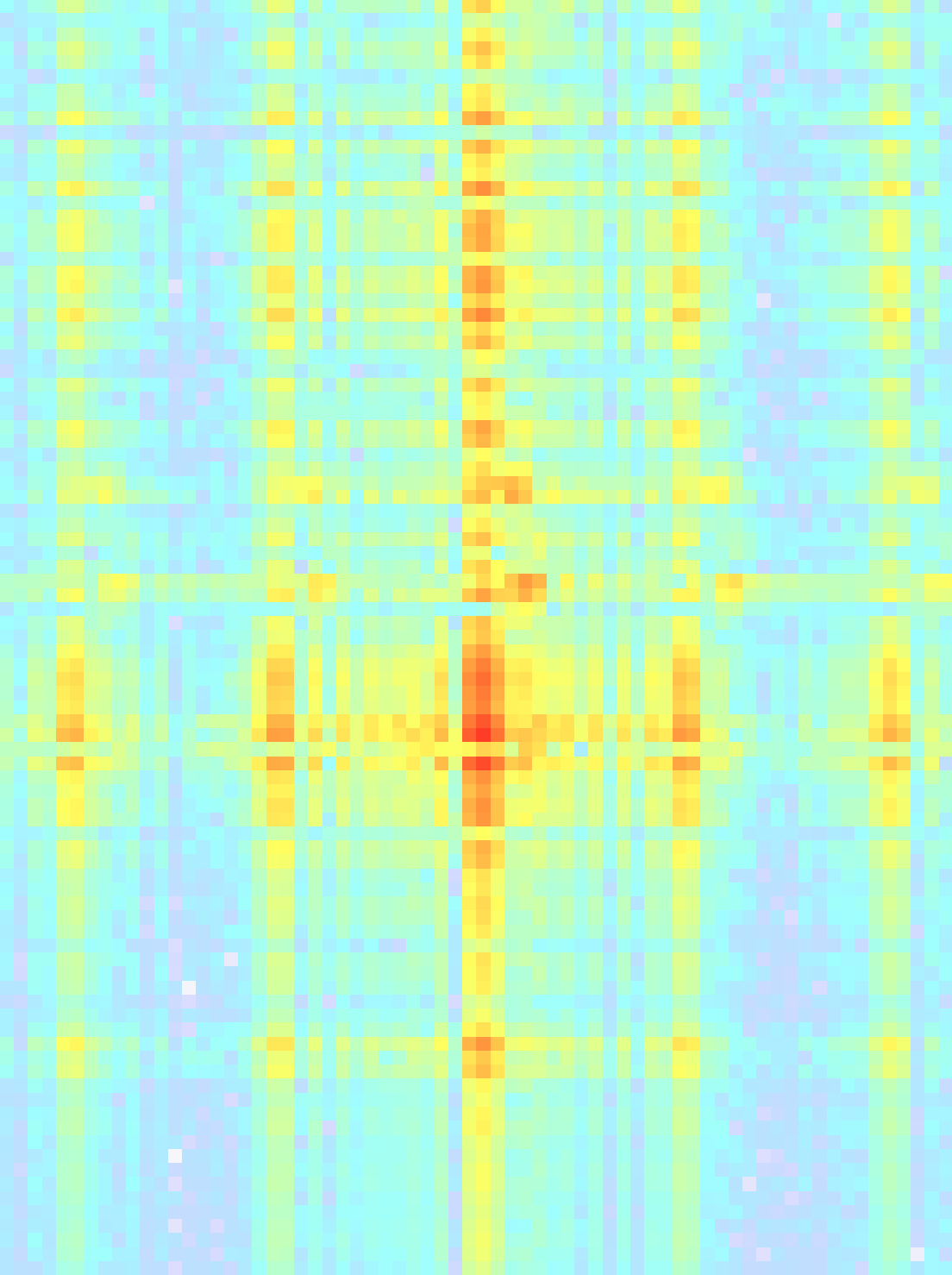};  %

            \addplot[only marks, mark=x, mark size=4pt, legend image post style={mark size=2pt}, legend image post style={mark size=4pt}] table[x expr=\thisrow{speed_mmps}/1000, y expr=\thisrow{range_mm}/1000, col sep=comma]{./tikz/data/postprocessing/object_list/objectlist_groundtruth_01353.dat};
            \label{plots:periodogram:groundtruth}

            \addplot[only marks, mark=o, mark size=2.5pt, mark options={fill=none, color=uni_rot, draw=uni_rot, line width=1pt}, legend image post style={mark size=3pt}] table[x expr=\thisrow{speed_mmps}/1000, y expr=\thisrow{range_mm}/1000, col sep=comma]{./tikz/data/postprocessing/object_list/objectlist_crap_01353.dat};
            \label{plots:periodogram:detected}

        \nextgroupplot[xlabel={Speed in \unit{\meter\per\second}\\(b)}, xlabel style={align=center}]

            \addplot graphics [xmin=-10, xmax=10, ymin=-0.3030, ymax=60.3030] {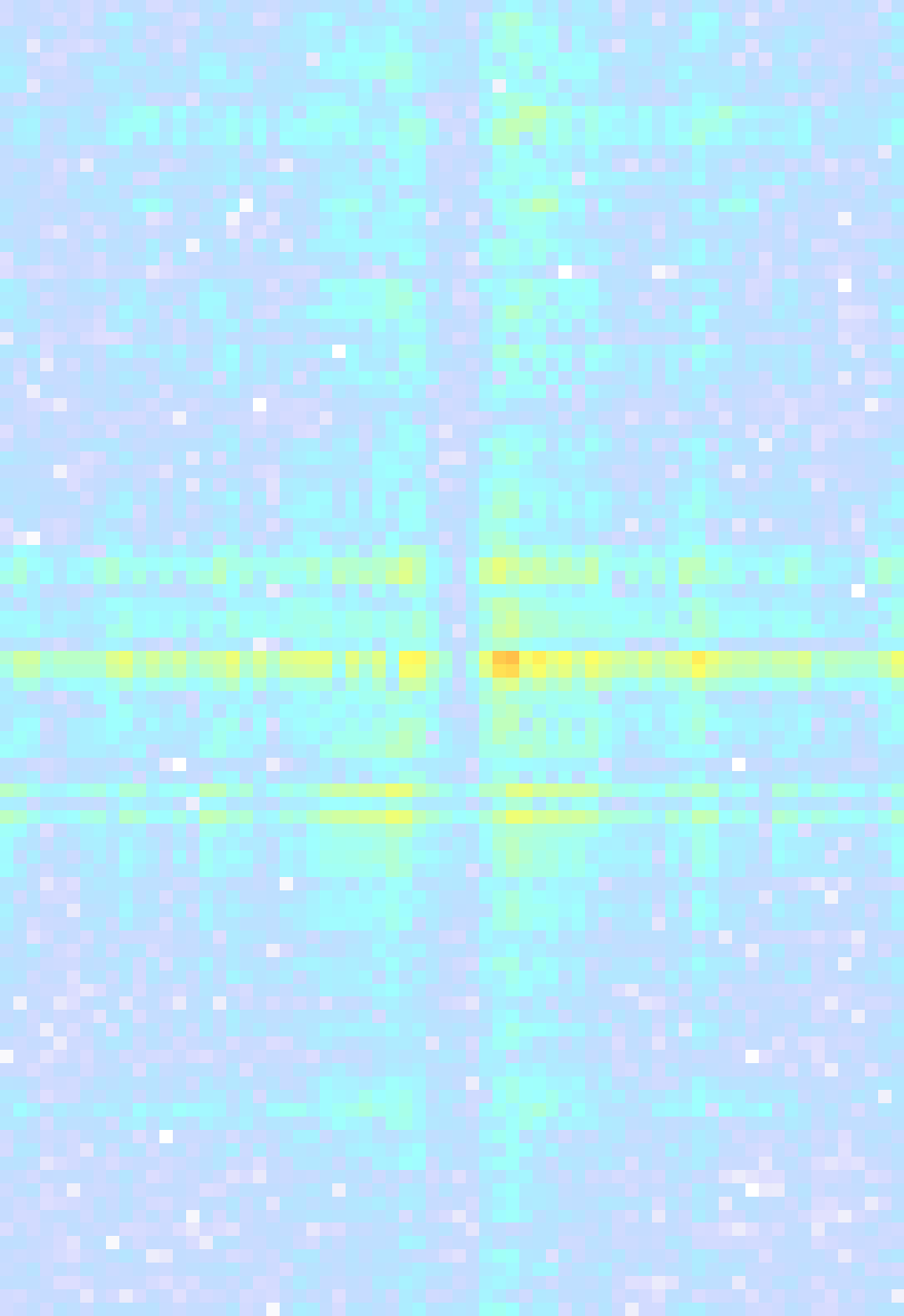};  %

            \addplot[only marks, mark=x, mark size=4pt, legend image post style={mark size=2pt}] table[x expr=\thisrow{speed_mmps}/1000, y expr=\thisrow{range_mm}/1000, col sep=comma]{./tikz/data/postprocessing/object_list/objectlist_groundtruth_01353.dat};

            \addplot[only marks, mark=o, mark size=2.5pt, mark options={fill=none, color=uni_rot, draw=uni_rot, line width=1pt}] table[x expr=\thisrow{speed_mmps}/1000, y expr=\thisrow{range_mm}/1000, col sep=comma]{./tikz/data/postprocessing/object_list/objectlist_ecac_01353.dat};

        \nextgroupplot[xlabel={Speed in \unit{\meter\per\second}\\(c)}, xlabel style={align=center}]

            \addplot graphics [xmin=-10, xmax=10, ymin=-0.3030, ymax=60.3030] {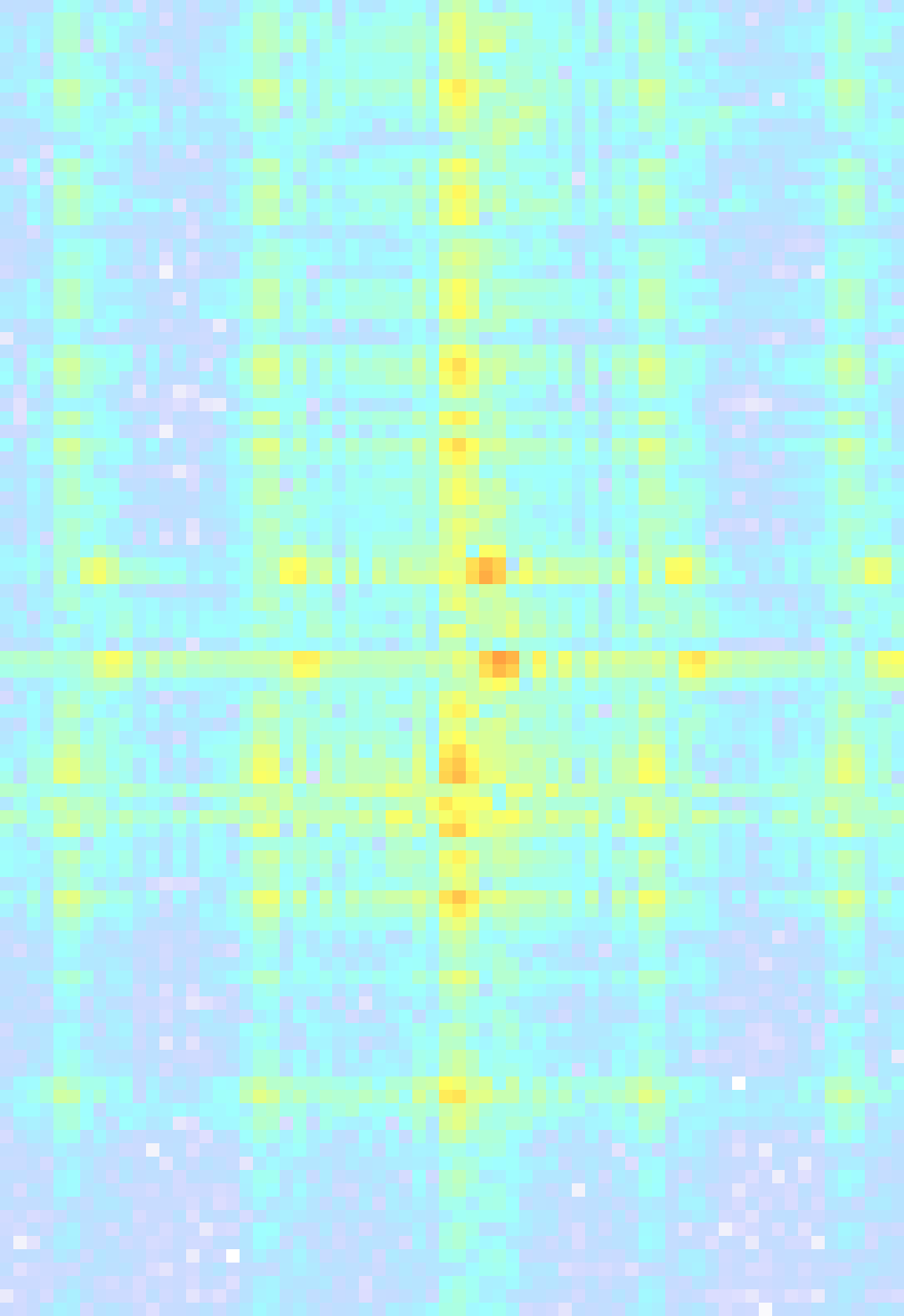};  %

            \addplot[only marks, mark=x, mark size=4pt, legend image post style={mark size=2pt}] table[x expr=\thisrow{speed_mmps}/1000, y expr=\thisrow{range_mm}/1000, col sep=comma]{./tikz/data/postprocessing/object_list/objectlist_groundtruth_01353.dat};

            \addplot[only marks, mark=o, mark size=2.5pt, mark options={fill=none, color=uni_rot, draw=uni_rot, line width=1pt}] table[x expr=\thisrow{speed_mmps}/1000, y expr=\thisrow{range_mm}/1000, col sep=comma]{./tikz/data/postprocessing/object_list/objectlist_selfcal2_01353.dat};

        \nextgroupplot[colorbar, colormap name=jet_inue, colorbar style={ylabel={Normalized power in \unit{\decibel}}, at={(1.05,0)}, anchor=south west, yticklabel style={font=\small}, ylabel style={yshift=3.5mm, font=\small}, ytick={-100,-80,...,0}, width=2mm}, point meta min=-110, point meta max=0, xlabel={Speed in \unit{\meter\per\second} \\ (d)}, xlabel style={align=center}]

            \addplot graphics [xmin=-10, xmax=10, ymin=-0.3030, ymax=60.3030] {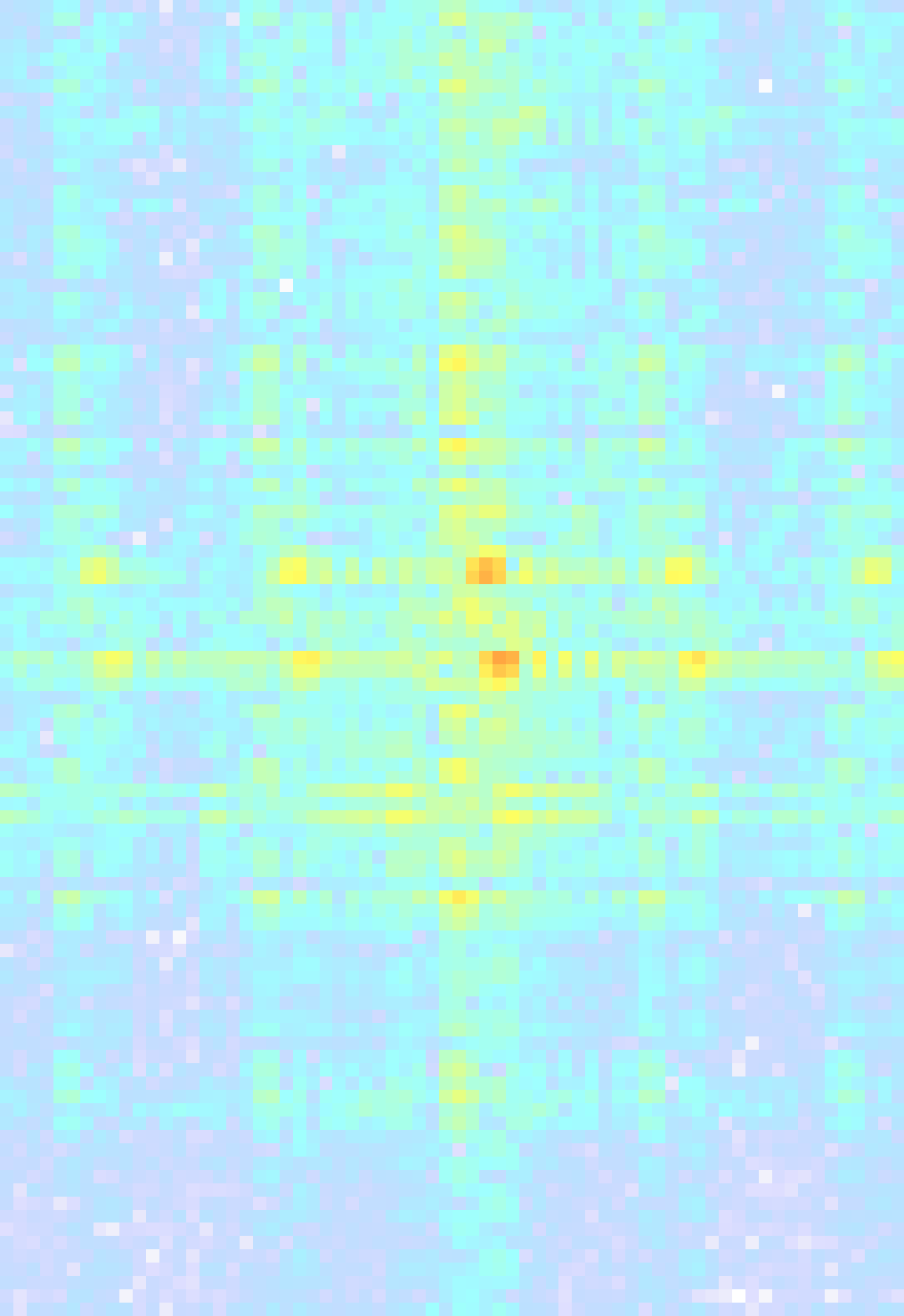};  %

            \addplot[only marks, mark=x, mark size=4pt, legend image post style={mark size=2pt}] table[x expr=\thisrow{speed_mmps}/1000, y expr=\thisrow{range_mm}/1000, col sep=comma]{./tikz/data/postprocessing/object_list/objectlist_groundtruth_01353.dat};

            \addplot[only marks, mark=o, mark size=2.5pt, mark options={fill=none, color=uni_rot, draw=uni_rot, line width=1pt}] table[x expr=\thisrow{speed_mmps}/1000, y expr=\thisrow{range_mm}/1000, col sep=comma]{./tikz/data/postprocessing/object_list/objectlist_rect_01353.dat};

	\end{groupplot}
\end{tikzpicture}
    \vspace*{-6mm}
    \caption{Examples of measured periodograms after different steps of the processing pipeline.
    The ground truth targets are marked with (\ref{plots:periodogram:groundtruth}) and the \acrshort{cfar}-detections with (\ref{plots:periodogram:detected}).
    (a) Original periodogram directly from the \acrshort{poc}, (b) after clutter removal with \acrshort{eca-c}, (c) after clutter removal with \acrshort{crap}, (d) final version after \acrshort{tdd} peak detection.}
	\label{fig:pipeline:periodograms}
    \vspace*{-3mm}
\end{figure*}

Fig.~\ref{fig:pipeline:blockdiagram} depicts our implemented processing pipeline processing the raw IQ samples into range and radial speed estimates, which are fed to the actual tracker instance.

We employ standard \gls{ofdm} \acrshort{radar} processing to compute the periodogram based on the IQ samples of the \gls{isac} \gls{poc}.
First, the received frame is divided element-wise by the transmitted frame to extract the \gls{csi} matrix $\boldsymbol{H}\kern-1pt \in \mathbb{C}^{N \times M}$ with $N$ subcarriers and $M$ \gls{ofdm} symbols.
After potentially applying windowing for sidelobe suppression, we compute \glspl{dft} over the symbols (slow time) of $\boldsymbol{H}$ for radial speed information and I\glspl{dft} over the subcarriers (fast time) of $\boldsymbol{H}$ for range estimates.
For more details about \gls{ofdm} \acrshort{radar} processing, please refer to \cite{braun_diss}.

Fig.~\ref{fig:pipeline:periodograms}a shows an example of a resulting raw periodogram directly from \gls{poc} data.
The detected peaks are marked with (\ref{plots:periodogram:detected}) and ground truth with (\ref{plots:periodogram:groundtruth}).
It is evident that there are a lot of false detections, which are explained below, along with ways to mitigate them. %

\vspace*{-1.5mm}
\subsection{Clutter Removal}\label{subsec:pipeline_clutter}
As a first step, clutter removal suppresses unwanted components in the raw periodogram.
Here, we have to deal with static reflecting components in the ARENA2036 environment, e.g., a big metal cargo gate. %
To mitigate the impact of clutter, we tested two different algorithms:

\begin{enumerate}
\item An adapted version of \gls{eca-c}~\cite{zhao_ecac}, modified as described in~\cite{henninger_crap}.
As can be seen in Fig.~\ref{fig:pipeline:periodograms}b, the downside of this approach is that it suppresses all zero Doppler component.
Thus, also potential targets with an absolute radial speed below the speed resolution of the system \mbox{($\approx \qty{0.56}{\meter\per\second}$)} are strongly attenuated, making the approach unsuitable for our indoor tracking use case.

\item \Gls{crap} \cite{henninger_crap}, leverages vectorization to perform clutter suppression in the time-frequency domain. Fig.~\ref{fig:pipeline:periodograms}c shows that this approach does not attenuate the static clutter components as strongly as \gls{eca-c}. However, it allows the detection of slow-moving targets, making it more suitable for our application.
Note that for the clutter acquisition phase of \gls{crap}, we used frames from the actual tracking experiments, i.e., not from dedicated reference measurements from an empty scenario.
\end{enumerate}

\vspace*{-2mm}
\subsection{TDD Peak Detection}\label{subsec:pipeline_tdd}
To extract peaks from the periodogram, we use the well-established \gls{cfar} algorithm.
Specifically, we employ cell-averaging (CA)-\gls{cfar} \cite{richards2010principles}, which is efficient with a suitable kernel and 2D convolution over the periodogram. The cell (or bin) under test is compared to the average power of the cells around it (reference cells); if it exceeds the threshold it is declared a peak.
Hereby, the immediately adjacent cells  (guard cells) are excluded, as they might contain contributions attributable to the cell under test.

However, the sole application of a \acrshort{cacfar} detector results in numerous false alarms due to periodic detections along the speed direction in addition to the true target peaks. This can be observed in Figs.~\ref{fig:pipeline:periodograms}a-c. Those wrong detections are due to the \gls{tdd} between periodic \gls{doli} and \gls{upli} transmissions in 3GPP-compliant communication. Since only reflections of the \gls{doli} signal are processed for sensing, the empty \gls{upli} symbols act as a time domain windowing to the sensing \gls{csi}.
This leads to weighted, periodic repetitions of the target peak (impulsive sidelobes), as described in detail in \cite{henninger2025targetdetectionisactdd}.
To circumvent this issue, we implement the technique from~\cite{henninger2025targetdetectionisactdd} with the core idea of using knowledge about the \gls{tdd} windowing pattern to compensate for it.
More specifically, an iterative routine to detect peaks in the periodogram is proposed. 
It consists of (i) focused Fourier analysis around the current candidate peak to obtain accurate range, Doppler shift, and complex coefficient estimates, (ii) reconstruction of the \gls{psf} (including the \gls{tdd} windowing effect) of the peak using the accurate estimates, and subtraction of the \gls{psf} from the original periodogram, and (iii) checking the power of the impulsive sidelobes after subtraction.
Contrary to peaks caused by impulsive sidelobes, removing the \gls{psf} of a true target peak will reveal a reduction in impulsive sidelobe power. As can be seen in Fig.~\ref{fig:pipeline:periodograms}d, this approach allows to only detect valid targets, while peaks due to \gls{tdd}-induced impulsive sidelobes are correctly rejected.

Finally, we filter the peak list to get the final list of estimates that are fed to the tracker.
This includes a priori knowledge, e.g., a scenario dependent maximum possible range, and an experimentally determined minimal peak power.

\section{Tracking}\label{sec:tracker}
We implemented a basic \gls{phd} filter from the \gls{rfs} framework for \gls{mtt}.
Specifically, we implemented its \gls{gmix} version \cite{vo_gmphd} as the \gls{phd} recursion has, in general, no closed-form solution.
This filter is an unlabeled tracker and bears some similarities to the standard \gls{kf}.
Although this tracker is state-of-the-art, it does not provide label data.
This means that from the target state at one point in time, no conclusions can be drawn about the previous state, i.e., no time-wise association.
Its main benefit over more basic versions of multi-target trackers -- for example \gls{gnn} or \gls{jpdaf} -- is that it can handle target death, birth, and spawning.

The basis of the \gls{phd} filter is the abstraction of the multi-target state set to an (target) intensity function $\nu$, giving the estimated target density for an arbitrary state $\boldsymbol{x}$ in the scenario.
This enables its propagation through the multi-target extension of the Bayes filter under the framework of \gls{fisst}~\cite{mahler_fisst}.
The \gls{gmix} implementation assumes a posterior intensity of that form, i.e., with $J$ components with weights $w^{(i)}$, means $\boldsymbol{m}^{(i)}$, and covariances $\boldsymbol{P}^{(i)}$, at time $k-1$
\begin{equation}
    \nu_{k-1}(\boldsymbol{x}) = \sum_{i=1}^{J_{k-1}}w_{k-1}^{(i)}\mathcal{N}\left( \boldsymbol{x}; \boldsymbol{m}_{k-1}^{(i)},\boldsymbol{P}_{k-1}^{(i)} \right)\,. %
\end{equation}
The predicted intensity is a combination of the predicted survival- $\nu_{S,k|k-1}(\boldsymbol{x})$, spawned- $\nu_{\beta,k|k-1}(\boldsymbol{x})$ and newly born components $\gamma_k(\boldsymbol{x})$
\begin{equation}
    \nu_{k|k-1}(\boldsymbol{x}) = \nu_{S,k|k-1}(\boldsymbol{x}) + \nu_{\beta,k|k-1}(\boldsymbol{x}) + \gamma_k(\boldsymbol{x})\,,
\end{equation}
with the probability of survival $p_{S,k}$, state transition matrix $\boldsymbol{F}_{k-1}$, process noise covariance $\boldsymbol{Q}_{k-1}$, and
\begin{multline}
    \nu_{S,k|k-1}(\boldsymbol{x}) = p_{S,k}\sum_{j=1}^{J_{k-1}}w_{k-1}^{(j)}\mathcal{N}\left( \boldsymbol{x};\boldsymbol{F}_{k-1}\boldsymbol{m}_{k-1}^{(j)},\right.\\
    \left.\boldsymbol{Q}_{k-1}+\boldsymbol{F}_{k-1}\boldsymbol{P}_{k-1}^{(j)}\boldsymbol{F}_{k-1}^\intercal \right) \,.
\end{multline}
After the update, the posterior intensity is then
\begin{equation}
    \nu_{k}(\boldsymbol{x}) = (1-p_{D,k})\nu_{k|k-1}(\boldsymbol{x}) + \sum_{\boldsymbol{z} \in Z_k}\nu_{D,k}(\boldsymbol{x};\boldsymbol{z})\,,
\end{equation}
including the probability of detection $p_{D,k}$, measurements $\boldsymbol{z} \in Z_k$, and the updated components $\nu_{D,k}(\boldsymbol{x};\boldsymbol{z})$.
It further includes pruning and merging of the \glspl{gmix} after each iteration.
More details can be found in \cite{vo_gmphd}.

For the tracker evaluation, we use simple comprehensive metrics.
While there are specific metrics, like OSPA\textsuperscript{(2)}~\cite{beard_ospa2}, designed for the peculiarities of \gls{mtt}, they are basically a combination of metrics for the individual issues -- such as track label switches, track interruption, etc. -- and are not necessary for our label-free problem at hand.
For the tracking accuracy, we use the \gls{mae} on both the range and radial speed components.
Hereby, the estimates are associated with the ground truth via the Hungarian Algorithm with a cutoff of \qty{5}{\meter} in range and \qty{5}{\meter\per\second} in speed.
Also the probability of detection is calculated based on the association and not just based on the number of detections.
Given our high update rate, we employ windowing over \qty{11} frames (\qty{110}{\milli\second}) for an target to be correctly marked as detected.
To evaluate clutter and false alarm suppression, we provide the average number of false alarms per scan.
For visualization, we provide the averaged estimated cardinality (or number of targets) over \qty{51} frames, which is a reasonable assumption given the \gls{phd} filter provides the correct number of targets only on average.

\section{Results}\label{sec:results}
In this section we present the scenarios, configuration parameters, as well as the tracking results both for the measurements and the simulated baseline.

\subsection{Scenarios}\label{subsec:results:sceanrios}
\begin{figure}[t]
	\centering
  	\begin{tikzpicture}
    \centering
    \begin{groupplot}[
        group style={group name=scenario, group size=1 by 3, vertical sep=5mm, horizontal sep=0mm},
        width=0.51\textwidth, height=3cm, %
        grid=both, grid style={solid,unigrau!20}, %
        xmin=-1, xmax=31, xtick={0,5,...,30}, xticklabels={,,}, xlabel={},
        xlabel near ticks, xlabel style={yshift=1mm}, minor x tick num=1,
        ylabel={},
        ylabel near ticks, ylabel style={at={(-0.07,0.5)}},
        ]

        \nextgroupplot[ymin=15, ymax=55, ytick={0,10,20,...,50}, yticklabels={0,10,20,...,50}, ylabel={\small{Range in \unit{\meter}}}, minor y tick num=1]

            \draw[uni_rot, solid] (12.17,0) -- (12.17,60);

            \addplot[color=uni_mittelblau, solid, line width=1pt] table[x expr=\thisrow{timestamp}/1000, y=range1, col sep=comma]{./tikz/data/groundtruth_scenarios/scenario_4_groundtruth.dat};
            \addplot[color=uni_apfelgruen, solid, line width=1pt] table[x expr=\thisrow{timestamp}/1000, y=range2, col sep=comma]{./tikz/data/groundtruth_scenarios/scenario_4_groundtruth.dat};
            \addplot[color=lavenderplot, solid, line width=1pt] table[x expr=\thisrow{timestamp}/1000, y=range3, col sep=comma]{./tikz/data/groundtruth_scenarios/scenario_4_groundtruth.dat};
            \addplot[color=uni_gelb, solid, line width=1pt] table[x expr=\thisrow{timestamp}/1000, y=range4, col sep=comma]{./tikz/data/groundtruth_scenarios/scenario_4_groundtruth.dat};
            \addplot[color=black, solid, line width=1pt] table[x expr=\thisrow{timestamp}/1000, y=range5, col sep=comma]{./tikz/data/groundtruth_scenarios/scenario_4_groundtruth.dat};
            \addplot[color=darkgray176, solid, line width=1pt] table[x expr=\thisrow{timestamp}/1000, y=range6, col sep=comma]{./tikz/data/groundtruth_scenarios/scenario_4_groundtruth.dat};

        \nextgroupplot[ymin=-5.5, ymax=5.5, ytick={-5,0,...,5}, ylabel={\small{Speed in \unit{\meter\per\second}}}, minor y tick num=4,]

            \draw[uni_rot, solid] (12.17,-10) -- (12.17,10);

            \addplot[color=uni_mittelblau, solid, line width=1pt] table[x expr=\thisrow{timestamp}/1000, y=doppler1, col sep=comma]{./tikz/data/groundtruth_scenarios/scenario_4_groundtruth.dat};
            \addplot[color=uni_apfelgruen, solid, line width=1pt] table[x expr=\thisrow{timestamp}/1000, y=doppler2, col sep=comma]{./tikz/data/groundtruth_scenarios/scenario_4_groundtruth.dat};
            \addplot[color=lavenderplot, solid, line width=1pt] table[x expr=\thisrow{timestamp}/1000, y=doppler3, col sep=comma]{./tikz/data/groundtruth_scenarios/scenario_4_groundtruth.dat};
            \addplot[color=uni_gelb, solid, line width=1pt] table[x expr=\thisrow{timestamp}/1000, y=doppler4, col sep=comma]{./tikz/data/groundtruth_scenarios/scenario_4_groundtruth.dat};
            \addplot[color=black, solid, line width=1pt] table[x expr=\thisrow{timestamp}/1000, y=doppler5, col sep=comma]{./tikz/data/groundtruth_scenarios/scenario_4_groundtruth.dat};
            \addplot[color=darkgray176, solid, line width=1pt] table[x expr=\thisrow{timestamp}/1000, y=doppler6, col sep=comma]{./tikz/data/groundtruth_scenarios/scenario_4_groundtruth.dat};

        \nextgroupplot[xlabel={Time in \unit{\second}}, xticklabels={0,5,...,30},
                        ymin=-0.5, ymax=6.5, ytick={0,2,...,6}, yticklabels={0,2,...,6}, ylabel={\small{Cardinality}}, minor y tick num=1]

            \draw[uni_rot, solid] (12.17,-10) -- (12.17,10);

            \addplot[color=black, solid, line width=1pt] (-1,0) -- (-0.01,0);
            \addplot[color=black, solid, line width=1pt] (0,1) -- (1.98,1);
            \addplot[color=black, solid, line width=1pt] (1.99,3) -- (3.98,3);
            \addplot[color=black, solid, line width=1pt] (3.99,4) -- (5.98,4);
            \addplot[color=black, solid, line width=1pt] (5.99,5) -- (9.99,5);
            \addplot[color=black, solid, line width=1pt] (10,6) -- (17.99,6);
            \addplot[color=black, solid, line width=1pt] (18,5) -- (23.99,5);
            \addplot[color=black, solid, line width=1pt] (24,4) -- (25.99,4);
            \addplot[color=black, solid, line width=1pt] (26,2) -- (28.99,2);
            \addplot[color=black, solid, line width=1pt] (29,1) -- (29.99,1);
            \addplot[color=black, solid, line width=1pt] (30,0) -- (31,0);
            
	\end{groupplot}
\end{tikzpicture}
    \vspace*{-5mm}
	\caption{Scenario 4 with six targets, used for target emulation and as ground truth for tracker evaluation.
    The range and radial speed components of the targets as well as their cardinality are plotted over the simulation time of \qty{30}{\second}.
    Targets are born and die at different times and their crossing is marked with a vertical red line.}
	\label{fig:results:scenarios}
\end{figure}
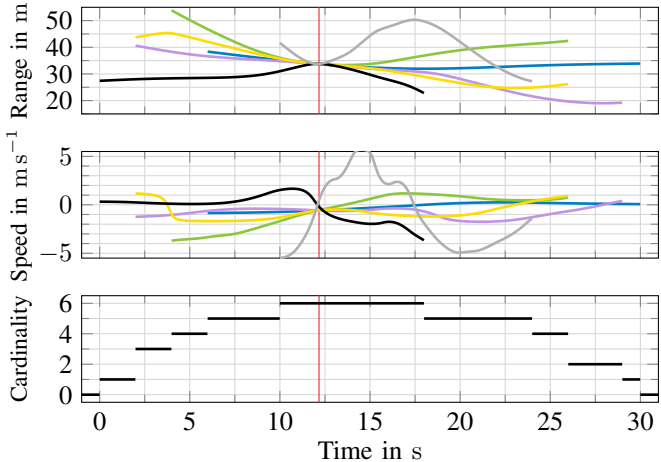

We evaluate four scenarios of $\qty{30}{\second}$ duration.
Their trajectories are generated via random walks of changing speeds with turn rate and maximum velocity constraints~\cite{bauhofer_multitarget}.
The scenarios resemble indoor traffic scenarios (\mbox{$\qty{18}{\meter} \le \text{range} \le \qty{54}{\meter}$}) with multiple pedestrians (absolute radial speed $\le \qty{5.6}{\meter\per\second}$, \gls{rcs} $\qty{1}{\meter\squared}$).
They should cover a broad range of possible real-world cases including target birth and death.\newline
Baseline scenario 1 is similar to scenario 2 with two targets but no crossing (i.e., equal states at a timestamp).
Scenario 2 is depicted in Fig.~\ref{fig:results:rd_plot} (left), with the two targets crossing around $\qty{8}{\second}$.
Scenario 3 in Fig.~\ref{fig:results:rd_plot} (right) are six targets with multiple crossings of up to 3 targets.
Scenario 4 in Fig.~\ref{fig:results:scenarios} features the same targets as Scenario 3, but adjusted and shifted to cross at a single point in time near $\qty{12}{\second}$, representing an extreme case.

\subsection{Configuration Parameters}\label{subsec:results:params}
Measurement setup, \gls{poc}, and target emulation are described in Sec.~\ref{sec:setup}.
Tab.~\ref{tab:observation:parameters} lists the parameters of the \gls{poc} as well as those for the pure baseline simulation for comparison.
In the latter, the real-world scenario with the \gls{poc} is replaced by a simulated range-Doppler sensor with \gls{awgn} noise running on the same scenarios.
These parameters should resemble the actual \gls{poc} performance~\cite{bauhofer_multitarget,mandelli_isacsurvey}.
The false alarms are uniform in range with zero Doppler, and their cardinality Poisson distributed.\newline
The processing pipeline uses no window function (i.e., a rectangular window) and filters detections outside the interval \mbox{[15,60]\,\unit{m}} and \mbox{[-6,6]\,\unit{\meter\per\second}} as well as below \qty{-40}{\decibel}.
Tab.~\ref{tab:tracker:parameters} provides the parameters we use for the \gls{phd} filter.
While these parameters are constant throughout all scenarios, some of them are optimized and set based on this environment experiments.

\begin{table}[t]
	\caption{Parameters for the observation generation}
    \label{tab:observation:parameters}
    \resizebox{\columnwidth}{!}{%
	
\begin{tabular}{lll}
    \toprule
    Domain & Parameter & Value \\
    \midrule
    \multirow{4}{*}{Measurements} & Update rate & \qty{100}{\hertz} \\
     & Frequency & \qty{27.6}{\giga\hertz} \\
     & Setup & (quasi-) monostatic \\
     & Waveform & \acrshort{ofdm} \\
     & Subcarrier spacing & \qty{120}{\kilo\hertz} \\
     & Number of subcarriers & $N=1584$ \\
     & Number of \gls{ofdm} symbols per frame & $M=1120$ (per \gls{tdd} pattern: 104 DL, 36 UL) \\
    \midrule
    \multirow{5}{*}{Simulations} & Update rate & \qty{100}{\hertz} \\
     & Noise variance range & $\sigma_\mathrm{r}^2=\qty{2.5e-5}{\meter\squared}$ \\
     & Noise variance radial speed & $\sigma_\mathrm{D}^2=\qty{7.396e-3}{\meter\squared\per\second\squared}$ \\
     & Probability of detection & $P_\text{det}=0.8$ \\
     & False alarms (Poisson distr., per scan) & $\lambda_\text{FA}=0.2$ \\
    \bottomrule
\end{tabular}

    }
\end{table}

\begin{table}[t]
	\caption{Parameters for the tracker}
	\label{tab:tracker:parameters}
    \resizebox{\columnwidth}{!}{%
	
\begin{tabular}{ll}
    \toprule
    Parameter & Value or Range \\
    \midrule
    Tracking domain & Range and radial speed \\
    Probability of survival & $p_{S,k} = p_{S} = 0.93$ \\
    Probability of detection & $p_{D,k} = p_{D} = 0.8$ \\
    State transition matrix & $\boldsymbol{F}_{k-1}(t) = \boldsymbol{F}(t) = \begin{bmatrix} 1 & t \\ 0 & 1 \end{bmatrix}$ \\
    Process noise covariance (wrt. \qty{1}{\second}) & $\boldsymbol{Q}_{k} = \boldsymbol{Q} = \begin{bmatrix} 0.01 & 0 \\ 0 & 0.01 \end{bmatrix}$ \\
    Spawn intensity & $\nu_\beta = 0$ \\
    Birth intensity & $\boldsymbol{\gamma}_k = \num{1e-6}\cdot\mathcal{N}\left( \boldsymbol{\mu}=\begin{bmatrix} 20 \\ 0.5 \end{bmatrix},\boldsymbol{\Sigma}=\begin{bmatrix} 100 & 0 \\ 0 & 100 \end{bmatrix} \right)$ \\
    \acrshort{gmix} merging radius (euclidean distance) & 0.4 \\
    \acrshort{gmix} pruning threshold & None \\
    \acrshort{gmix} pruning max number of components & 40 \\
    \bottomrule
\end{tabular}

    }
    \vspace*{-3mm}
\end{table}

\subsection{Evaluation}\label{subsec:results:evaluation}
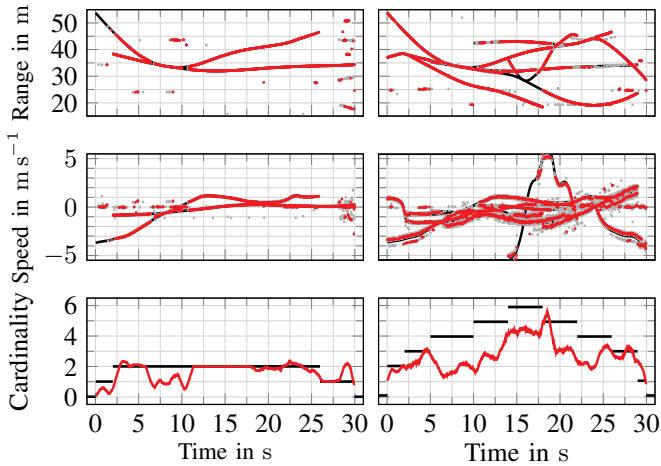
\begin{figure}[t!]
    \centering
  	\begin{tikzpicture}
    \centering
    \begin{groupplot}[
        group style={group size=2 by 3, vertical sep=5mm, horizontal sep=2mm},
        width=0.29\textwidth, height=3cm, %
        grid=both, grid style={solid,unigrau!20}, %
        xmin=-1, xmax=31, xtick={0,5,...,30}, xticklabels={,,}, xlabel={},
        xlabel near ticks, xlabel style={yshift=1mm}, minor x tick num=1,
        ylabel={},
        ylabel near ticks, ylabel style={at={(-0.16,0.5)}},
        ]

        \nextgroupplot[ymin=15, ymax=55, ytick={0,10,20,...,50}, yticklabels={0,10,20,...,50}, ylabel={\small{Range in \unit{\meter}}}, minor y tick num=1]

            \addplot graphics [xmin=-1,xmax=31, ymin=15,ymax=55,] {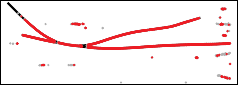};

            \addplot[color=black, solid, line width=1pt] (-10,10);
            \label{plots:results_rd_plot_gt}
            \addplot[only marks, mark=*, mark options={fill=darkgray176, color=darkgray176}, mark size=0.3pt, legend image post style={mark size=1.5pt}] (-10,10);
            \label{plots:results_rd_plot_obs}
            \addplot[only marks, mark=*, mark options={fill=uni_rot, color=uni_rot}, mark size=0.3pt, legend image post style={mark size=1.5pt}] (-10,10);
            \label{plots:results_rd_plot_est}

        \nextgroupplot[ymin=15, ymax=55, ytick={0,10,20,...,50}, yticklabels={,,}, minor y tick num=1]

            \addplot graphics [xmin=-1,xmax=31, ymin=15,ymax=55] {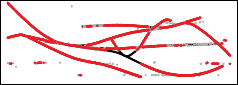};

        \nextgroupplot[ymin=-5.5, ymax=5.5, ytick={-5,0,...,5}, ylabel={\small{Speed in \unit{\meter\per\second}}}, minor y tick num=4]

            \addplot graphics [xmin=-1,xmax=31, ymin=-5.5,ymax=5.5] {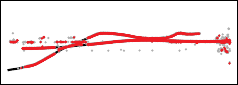};
        
        \nextgroupplot[ymin=-5.5, ymax=5.5, yticklabels={,,}, minor y tick num=4]

            \addplot graphics [xmin=-1,xmax=31, ymin=-5.5,ymax=5.5] {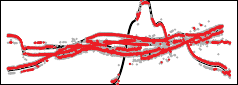};

        \nextgroupplot[xlabel={\small{Time in \unit{\second}}}, xticklabels={0,5,...,30},
                        ymin=-0.5, ymax=6.5, ytick={0,2,...,6}, yticklabels={0,2,...,6}, ylabel={Cardinality}, minor y tick num=1]

            \addplot graphics [xmin=-1,xmax=31, ymin=-0.5,ymax=6.5] {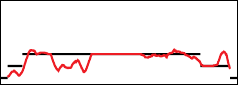};

        \nextgroupplot[xlabel={Time in \unit{\second}}, xticklabels={0,5,...,30},
                        ymin=-0.5, ymax=6.5, ytick={0,2,...,6}, yticklabels={,,}, minor y tick num=1]

            \addplot graphics [xmin=-1,xmax=31, ymin=-0.5,ymax=6.5] {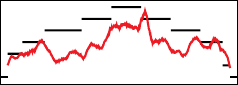};

	\end{groupplot}
\end{tikzpicture}
    \vspace*{-4mm}
	\caption{Tracking performance of the \gls{phd} filter with measurements for scenario 2 \textit{(left)} and scenario 3 \textit{(right)}.
    The ground truth is marked with (\ref{plots:results_rd_plot_gt}), (\ref{plots:results_rd_plot_obs}) are the observations fed to the tracker and (\ref{plots:results_rd_plot_est}) the actual tracker estimates.}
	\label{fig:results:rd_plot}
\end{figure}

Tab.~\ref{tab:results:performance} shows the tracking performance for the different scenarios, evaluating the simulations and the measurements.\newline
For the simulations, the range and radial speed \glspl{mae} are increasing with more demanding scenarios.
While the resolution impact is not considered in the simulated range-Doppler sensor, intersections of tracks yield higher \glspl{mae} as the merging stage of the \gls{phd} will collapse significant close states.
Nonetheless, this cannot be observed in the windowed probability of detection, which is high throughout all scenarios.
False alarms are effectively reduced.\newline
In comparison, for the measurements -- exemplarily plotted in Fig.~\ref{fig:results:rd_plot} for scenario 2 and 3 -- we observe worse performance.
The resolution limitations of the system are reflected in higher \gls{mae} values.
This is generally the case in these demanding scenarios, where the radial speeds are so small that they can rarely be used to separate targets close in range.
The \gls{phd} filter is not able to solve all these cases and just drops states in those areas.
Overall, the radial speed estimation performance is better due to its much smaller range of values.
For the range, large outliers from false alarms in combination with the birth process of the \gls{phd} drop its performance in the worst case to an \gls{mae} of up to \qty{1.4}{\meter}.
Also path loss is an issue here, which can be seen in the first seconds of scenario 2; no observations are available for the tracker at far distance due to the preset power filter.
The false alarms per scan are fairly low thanks to good settings in the processing pipeline and rejection from the filter itself.
The probability of detection drops to \qty{91}{\percent} for reasonable scenarios and clearly benefits from the high update rate of the system, rendering it adequate for a large number of use cases.
The underestimation of the cardinality is a result of the \gls{gmix} merging and a known property of the \gls{phd} filter~\cite{vo_cphd} for higher number of targets.
This further worsens the correct selection of estimates, even though the actual components are present in the \gls{gmix}-density, which can be observed by interrupted tracks.
For scenario 3, we can see an artifact of mirror Doppler observations, roughly \qty{0.5}{\meter\per\second} from actual targets.
This is probably a result of the digital target generation, which will inevitably result in sidelobes in the periodogram and possibly ghost targets due to the limited bandwidth.
While we see a notable discrepancy between simulations and measurements, we found they are mostly attributable to the infinite resolution in the former.
As a result, the simulations deviate too much from reality for a direct comparison, but are well suited for establishing an upper bound and confirming a properly working setup.

\begin{table}[t]
	\caption{Multi-target tracking performance}
	\label{tab:results:performance}
    \resizebox{\columnwidth}{!}{%
	
\begin{tabular}{llll}
    \toprule
    Scenario & Quantity/Metric & Simulated & Measured \\
    \midrule
    \multirow{4}{*}{Scenario 1} & Range \acrshort{mae} in \unit{\meter} & 0.01 & 0.05 \\
     & Speed \acrshort{mae} in \unit{\meter\per\second} & 0.02 & 0.03 \\
     & False alarms & \textless 0.01 & \textless 0.01 \\
     & Probability of detection & \qty{100}{\percent} & \qty{92.3}{\percent} \\
    \midrule
    \multirow{4}{*}{Scenario 2} & Range \acrshort{mae} in \unit{\meter} & 0.01 & 0.65 \\
     & Speed \acrshort{mae} in \unit{\meter\per\second} & 0.02 & 0.09 \\
     & False alarms & \textless 0.01 & 0.08 \\
     & Probability of detection & \qty{99.9}{\percent} & \qty{91.9}{\percent} \\
     \midrule
     \multirow{4}{*}{Scenario 3} & Range \acrshort{mae} in \unit{\meter} & 0.18 & 1.43 \\
     & Speed \acrshort{mae} in \unit{\meter\per\second} & 0.07 & 0.31 \\
     & False alarms & 0.04 & 0.12 \\
     & Probability of detection & \qty{99.3}{\percent} & \qty{91.1}{\percent} \\
     \midrule
     \multirow{4}{*}{Scenario 4} & Range \acrshort{mae} in \unit{\meter} & 0.13 & 0.69 \\
     & Speed \acrshort{mae} in \unit{\meter\per\second} & 0.08 & 0.08 \\
     & False alarms & 0.07 & 0.16 \\
     & Probability of detection & \qty{98.9}{\percent} & \qty{80.5}{\percent} \\
    \bottomrule
\end{tabular}

    }
    \vspace*{-5mm}
\end{table}

\section{Conclusion}\label{sec:conclusion}

We showcased the world's first \gls{isac} tracking with a real communication system.
We detailed our processing framework, from the acquisition to the final tracking results.
Given the challenging scenarios with multiple emulated pedestrian targets in a real-world industrial environment, our evaluation shows promising results for tracking in the range-Doppler domain with a state-of-the-art \gls{phd} filter.

In the future, we plan to conduct measurements with real targets, include the angular domain, improve and extend the tracking algorithm to include labels, and fuse multiple sensors.

\section*{Acknowledgments}

The authors would like to thank Artjom Grudnitsky for his valuable feedback and help during measurements.

This work was developed within the KOMSENS-6G project, partly funded by the German Ministry of Education and Research under grant 16KISK112K.

\balance

\bibliography{references}
\bibliographystyle{IEEEtran}

\end{document}